\documentclass[11pt]{article}

\usepackage{amsmath,amssymb}
\usepackage{graphicx}% Include figure files
\usepackage{caption}
\usepackage{color}% Include colors for document elements
\usepackage{dcolumn}% Align table columns on decimal point
\usepackage{bm}% bold math
\usepackage[numbers,super,comma,sort&compress]{natbib}

\usepackage{comment}
\usepackage{xy}
\xyoption{matrix}
\xyoption{frame}
\xyoption{arrow}
\xyoption{arc}

\usepackage{ifpdf}
\ifpdf
\else
\PackageWarningNoLine{Qcircuit}{Qcircuit is loading in Postscript mode.  The Xy-pic options ps and dvips will be loaded.  If you wish to use other Postscript drivers for Xy-pic, you must modify the code in Qcircuit.tex}
\xyoption{ps}
\xyoption{dvips}
\fi

\entrymodifiers={!C\entrybox}

\newcommand{\ket}[1]{{\left\vert{#1}\right\rangle}}
\newcommand{\qw}[1][-1]{\ar @{-} [0,#1]}
\newcommand{\qwx}[1][-1]{\ar @{-} [#1,0]}
\newcommand{\controlo}{*+<.01em>{\xy -<.095em>*\xycircle<.19em>{} \endxy}}

\newcommand{\ctrlo}[1]{\controlo \qwx[#1] \qw}
\newcommand{\multigate}[2]{*+<1em,.9em>{\hphantom{#2}} \POS [0,0]="i",[0,0].[#1,0]="e",!C *{#2},"e"+UR;"e"+UL **\dir{-};"e"+DL **\dir{-};"e"+DR **\dir{-};"e"+UR **\dir{-},"i" \qw}
\newcommand{\ghost}[1]{*+<1em,.9em>{\hphantom{#1}} \qw}
\newcommand{\push}[1]{*{#1}}
\newcommand{\gategroup}[6]{\POS"#1,#2"."#3,#2"."#1,#4"."#3,#4"!C*+<#5>\frm{#6}}
\newcommand{\rstick}[1]{*!L!<-.5em,0em>=<0em>{#1}}
\newcommand{\lstick}[1]{*!R!<.5em,0em>=<0em>{#1}}
\newcommand{\Qcircuit}{\xymatrix @*=<0em>}
\begin{document}

\title{Quantum chemistry beyond Born-Oppenheimer approximation on a quantum computer: a simulated phase estimation study}

\author{Libor Veis\footnote{These authors contributed equally.} $^{,}$\thanks{J. Heyrovsk\'{y} Institute of Physical Chemistry, Academy of Sciences of the Czech \mbox{Republic, v.v.i.}, Dolej\v{s}kova 3, 18223 Prague 8,
Czech Republic}, Jakub Vi\v{s}\v{n}\'ak$^{*,\dagger}$,
Hiroaki Nishizawa\thanks{Present address: Institute for Molecular Science, 38 Nishigo-Naka, Myodaiji, Okazaki,
444-8585, Japan}, Hiromi Nakai\thanks{Department of Chemistry and Biochemistry, School of Advanced Science and Engineering,
Waseda University, 3-4-1 Okubo, Shinjuku, Tokyo 169-8555, Japan}, Ji\v{r}\'{i} Pittner$^{\dagger}$
}

\date{\today}

\maketitle

\begin{abstract}

% LV: copmletely changed
We present an \textit{efficient} quantum algorithm for beyond-Born-Oppenheimer molecular energy computations. Our approach combines the quantum full 
configuration interaction method with the nuclear orbital plus molecular orbital (NOMO) method. We give the details of the algorithm and demonstrate 
its performance by classical simulations. Two isotopomers of the hydrogen molecule (H$_2$, HT) were chosen as representative examples and calculations 
of the lowest rotationless vibrational transition energies were simulated.

\end{abstract}

% LV: originally - too much
%\keywords{quantum computing, phase estimation, nuclear orbital plus molecular orbital method, Born-Oppenheimer approximation}

\bibliographystyle{apsrev}
\clearpage

\section{Introduction}

Exact computations and simulations of quantum systems on a classical computer are computationally hard. This stems from the fact that the dimensionality 
of the Hilbert space needed for the description of a studied quantum system scales exponentially with its size. One of the consequences is e.g. the 
prohibitive exponential scaling of the full configuration interaction (FCI) method.
Quantum computers \cite{nielsen_chuang}, on the other hand, offer an exponential speed-up for this task \cite{lloyd_1996, zalka_1998, ortiz_2001, somma_2002, abrams_1997, abrams_1999, ovrum_2007}, as was first noticed by Feynman and Manin \cite{feynman_1982, manin_1980}. The underlying idea, which employes 
mapping of the Hilbert space of a studied quantum system onto the Hilbert space of a register of quantum bits (qubits), both of them being exponentially, 
large, can in fact be adopted also in quantum chemistry. 

The past few years have witnessed a remarkable interest in the application of quantum computing for solving of different problems in quantum chemistry. 
Among others, quantum algorithms for non-relativistic \cite{aspuru-guzik_2005, wang_2008, whitfield_2010, veis_2010} as well as relativistic \cite{veis_2012}
 molecular FCI energy calculations, quantum chemical dynamics \cite{kassal_2008}, or calculations of molecular properties \cite{kassal_2009} were developed.
For a complete list of relevant papers, we refer the reader to recent reviews \cite{kassal_review, yung_review, veis_2012_chapter}.
\textit{Efficient} quantum chemical simulations are indeed believed to belong to the first practical applications of quantum computers. This is also 
supported by recent proof-of-principle few-qubit experiments \cite{lanyon_2010, du_2010, li_2011, lu_2011, lanyon_2011, peruzzo_2013}. Several improvements 
reducing the resource requirements of fault-tolerant implementation and thus paving the way for practical simulations were presented in \cite{jones_2012}. 

In this paper, motivated by the fact that non-Born-Oppenheimer (non-BOA) effects play an essential role in wide range fields (e.g. proton tunnelling in DNA 
damage), we generalise the applicability of the quantum FCI algorithm \cite{aspuru-guzik_2005, whitfield_2010} for beyond-BOA computations. 
We achieve this by combining qFCI with the NOMO method \cite{tachikawa_1998, nakai_2002, nakai_2003, nakai_2007,nomo-new1,nomo-new2,nomo-new3}. 
We should however note that our attempt is not the first one dealing with 
beyond-BOA computations on a quantum computer. In \cite{kassal_2008}, it was shown that simulating all electron-nuclear and inter-electronic interactions 
(and thus going beyond BOA) is somewhat surprisingly faster and more efficient than BOA for systems with more than four atoms. Nevertheless, 
the aforementioned approach is based on the first quantized formulation, thus completely different from ours.

The structure of the paper is following. In Section \ref{section_qfci} we shortly review the basic concepts of the phase estimation-based quantum FCI algorithm \cite{aspuru-guzik_2005, whitfield_2010}, in Section \ref{section_nomo} we do the same for the NOMO method, and in Section \ref{section_nomo_qfci} we present details of our quantum algorithm for beyond-BOA computations, which is a combination of both approaches. The performance of the proposed scheme is presented in Section \ref{section_application} by classical simulations of H$_2$ and HT energy computations.

\section{Quantum FCI algorithm}
\label{section_qfci}

An \textit{efficient} quantum FCI (qFCI) algorithm for calculations of nonrelativistic molecular energies employing the phase estimation algorithm (PEA) of Abrams and Lloyd \cite{abrams_1999} was proposed in the pioneering work by Aspuru-Guzik \textit{et al.} \cite{aspuru-guzik_2005}. It was later simplified by replacing of PEA with its iterative version, iterative phase estimation algorithm (IPEA) \cite{wang_2008, whitfield_2010, veis_2010}. (I)PEA is a quantum algorithm for obtaining the eigenvalue of a unitary operator $\hat{U}$, based on a given initial guess of the corresponding eigenvector. Since a unitary $\hat{U}$ can be written as $\hat{U} = e^{i\hat{H}}$, with $\hat{H}$ Hermitian, the (I)PEA can be viewed as a quantum substitute of the classical diagonalization.

Suppose that $| u \rangle$ is an eigenvector of $\hat{U}$ and that it holds

\begin{equation}
 \hat{U} | u \rangle = e^{2\pi i \phi} | u \rangle, \qquad \phi \in \langle 0, 1),
\end{equation}

\noindent
where $\phi$ is the phase which is estimated by the algorithm. In case of the original PEA, a quantum register is divided into two parts. The first one is the read-out part composed of $m$ qubits on which the binary representation of the estimate of phase $\phi$ is eventually measured. After the application of Hadamard gates in the read-out part of the quantum register followed by application of a sequence of controlled-$\hat{U}^{2^{k-1}}$ operations ($k$s are nonnegative integers from 1 to $m$), the register is transformed into

\begin{equation}
| \mathrm{reg} \rangle = \frac{1}{\sqrt{2^{m}}} \sum_{j=0}^{2^{m}-1} \hat{U}^{j} | j \rangle | u \rangle = \frac{1}{\sqrt{2^{m}}} \sum_{j=0}^{2^{m}-1} e^{2\pi ij \phi} | j \rangle | u \rangle. 
 \label{after_cu} 
\end{equation}

\noindent
The next part of the algorithm is the inverse quantum Fourier transform (QFT) \cite{nielsen_chuang} performed on the read-out part of the register. The whole register is transformed into $| 2^{m}\phi \rangle | u \rangle$ and the phase can be extracted from its first part.

\begin{figure}[!ht]
 %\begin{center}
 %\hskip 1cm
 %\mbox{
 %  \Qcircuit @C=1em @R=1em {
 %    \lstick{\ket{0}} & \gate{H} & \ctrl{1} & \gate{R_{z}(\omega_{k})} & \gate{H} & \meter & \cw & \phi_{k} \\
 %    \lstick{\ket{\psi_{\rm{system}}}} & {/} \qw & \gate{U^{2^{k-1}}} & {/} \qw & \qw
 %  }
 %}
 %\end{center}
	\begin{center}
    \includegraphics[width=8cm]{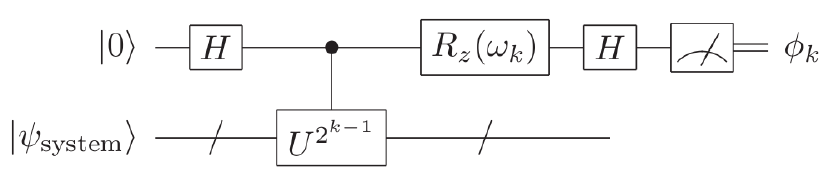}
 		\caption{
 \label{ipea_iteration}
The $k$-th iteration of the iterative phase estimation algorithm (IPEA). $H$ denotes Hadamard gate ($\pi/2$ rotation) and the feedback angle $\omega_k$ depends on the previously measured bits (see Eq. \ref{omega_k}).}
	\end{center}
\end{figure}

Iterative version, IPEA adopts the ideas of measurement-based quantum computing \cite{griffiths_1996} and reduces the computational resources by using only single read-out qubit. If $\phi$ is expressed in the binary form: $\phi = 0.\phi_1\phi_2\ldots\phi_m$, $\phi_i = \{ 0 , 1 \}$, one bit of $\phi$ is measured on the read-out qubit at each iteration step. The algorithm is iterated backwards from the least significant bits of $\phi$ to the most significant ones. The $k$-th iteration is shown in Figure \ref{ipea_iteration}.

\noindent
The equivalent of QFT is a single qubit $z$-rotation $R_{z}$, whose angle $\omega_{k}$ depends on the results of the previously measured bits

\begin{eqnarray}
 R_{z}(\omega_{k}) & = & \left( \begin{array}{cc} 
                            1 & 0 \\
                            0 & e^{2 \pi i \omega_{k}} \\
                            \end{array} \right) \\              
 \omega_{k} & = & -\sum_{i=2}^{m-k+1} \frac{\phi_{k+i-1}}{2^{i}}, \label{omega_k}
\end{eqnarray}

\noindent
followed by a Hadamard gate.

Depending on how the second part of the quantum register is treated in between individual iterations, we distinguish two versions of IPEA \cite{veis_2010}. In case of version \textbf{A}, it is maintained during all iterations (initialised only once). The biggest advantage of this approach is that one always ends up with one of the eigenstates of $\hat{U}$. On the other hand, the quantum coherence is needed for the whole algorithm which makes this version more difficult for an experimental realization. IPEA version \textbf{B} is on contrary characterised by reinitialization of the second part of the register at every iteration step. Therefore, the quantum coherence is needed only within each iteration separately.

The (I)PEA algorithm can be exploited for \textit{ab initio} quantum chemical calculations, if we take $\hat{U}$ in the form \cite{abrams_1999, aspuru-guzik_2005}

\begin{equation}
%  \hat{U} = e^{\displaystyle i\tau\hat{H}}
 \hat{U} = e^{i\tau\hat{H}},  
\end{equation}

\noindent
where $\hat{H}$ is the molecular electronic Hamiltonian (up to now, only Born-Oppenheimer Hamiltonians have been considered) and $\tau$ is a suitable parameter which assures $\phi$ being in the interval $\langle 0,1)$.

The electronic Hamiltonian can be expressed in the second quantized form as \cite{szabo_ostlund}

\begin{equation}
 \hat{H} = \sum_{pq} h_{pq} \hat{a}_{p}^{\dagger} \hat{a}_{q} + \frac{1}{2} \sum_{pqrs} V_{pqrs} \hat{a}_{p}^{\dagger} \hat{a}_{q}^{\dagger} \hat{a}_{s} \hat{a}_{r} = \sum_{X=1}^{L} \hat{h}_{X},
  \label{ham_sec_quant}
\end{equation}

\noindent
where $h_{pq}$ and $V_{pqrs}$ are one and two-electron integrals in the molecular spin orbital basis and $\hat{a}_{i}^{\dagger}$ and $\hat{a}_{i}$ are fermionic creation and annihilation operators. Since these operators in general do not commute, exponential of the Hamiltonian cannot be written as a product of exponentials of individual $\hat{h}_{X}$, but a numerical approximation must be used \cite{lloyd_1996}. The first-order Trotter approximation \cite{trotter} has the form

\begin{equation}
 e^{i\tau\hat{H}} = e^{i\tau\sum_{X=1}^{L}\hat{h}_{X}} = \Big(\prod_{X=1}^{L} e^{i\hat{h}_{X}\tau/N}\Big)^{N} + \mathcal{O}(\tau^{2}/N).
 \label{trotter}
\end{equation}

When representing the quantum chemical wave function on a quantum register, the simplest approach (but the least economical one in terms of number of qubits) is so called direct mapping. It directly assigns individual spin orbitals (or in relativistic case Kramers pair bispinors) to qubits, since they can be either occupied or unoccupied (occupation number basis), corresponding to $\ket{1}$ or $\ket{0}$ states. Jordan-Wigner transformation (JWT) \cite{jordan_1928} is then used to express fermionic operators in terms of Pauli $\sigma$ matrices. JWT has the form

\begin{equation}
\label{jordan-wigner}
 \hat{a}_{n}^{\dagger} = \Bigg( \bigotimes_{j=1}^{n-1} \sigma_{z}^{j} \Bigg) \otimes \sigma_{-}^{n}, \quad \hat{a}_{n} = \Bigg( \bigotimes_{j=1}^{n-1} \sigma_{z}^{j} \Bigg) \otimes \sigma_{+}^{n},
\end{equation}

\noindent
where $\sigma_{\pm} = 1/2(\sigma_{x} \pm i \sigma_{y})$ and the superscript denotes the qubit on which the matrix operates. Alternatively, the Bravyi-Kitaev transformation \cite{bravyi_2002, seeley_2012} which balances locality of occupation and parity information and reduces the simulation cost from $\mathcal{O}(n)$ to $\mathcal{O}(\log n)$ for one fermionic operator when compared to JWT may be used. We would like to note that more compact mappings e.g. from subspace of fixed-electron-number wave functions or spin-adapted wave functions can also be used \textit{efficiently} in connection with quantum sparse simulation algorithms \cite{toloui_2014}.

Regarding the overall scaling of the qFCI algorithm, Wecker \textit{et al.} \cite{wecker_2013} found that the computational time for bounded error scales with the number of spin orbitals $N$ as $\mathcal{O}(N^9)$ on average and as $\mathcal{O}(N^{11})$ at worst. Using the Bravyi-Kitaev transformation \cite{bravyi_2002, seeley_2012} instead of JWT would decrease the scaling to $\mathcal{O}(N^8 \log{N})$ or $\mathcal{O}(N^{10} \log{N})$ respectively. 
Poulin \textit{et al.} \cite{poulin_2014} used testing set of real molecules and observed even more feasible scaling of the Trotter-Suzuki time step leading to the overall scaling $\mathcal{O}(N^{4-5.5})$. The computational cost bounds can be further improved considering local molecular basis sets \cite{mcclean_2014}.

%Previously mentioned computational cost bounds can be further improved considering local molecular basis sets (e.g. symmetricaly orthogonalized atomic orbitals (OAO)) and constant fraction of filled spin-orbitals f = N_e/N (where N_e is the number of electrons). That leads to O(N) non-negligible one-particle and O(N^2) two-particle terms in hamiltonian (and O(N log N) and O(N^2 log N) terms in one Trotter-Suzuki time-step) and therefore to the computational time scaling as O(N^5 log N) [X2].
	
%It is also possible to exploit the fact that number of electron is a good quantum number for the electronic structure problem and therefore it is unnecessary to store the full Fock space of the orbitals [X4] - then the number of qubits needed scales as O(log_2 C(N,N_e)) and number of gates needed for one Trotter-Suzuki time-step as O((N-N_e)^{4/3}log log (N-N_e)) in general case and O((N-N_e)[log(N-Ne)]^{7/4}) using black-box algorithms with the best scaling with sparsity developed in [X5] (assuming equal norm terms).

The qFCI algorithm \cite{aspuru-guzik_2005, whitfield_2010} requires an initial guess of the exact eigenstate, whose quality influences the success probability of measuring the desired energy. This can be either a classical approximation [e.g. complete active space (CAS) based wave function \cite{wang_2008, veis_2010}], an exact state prepared by the adiabatic state preparation method \cite{aspuru-guzik_2005,veis_2014}, or by the algorithmic cooling method \cite{xu_2012}, or also a unitary coupled cluster approximation optimised by the recently presented combined classical-quantum variational approach \cite{peruzzo_2013, yung_2013}.

In order to increase the overall success probability of (I)PEA, the whole algorithm is repeated and the correct phase $\phi$ decided from the majority voting. In case of version \textbf{B}, individual iterations are independently repeated and correct $\phi_k$ values decided from the majority voting.

\section{NOMO method}
\label{section_nomo}

The nuclear orbital plus molecular orbital (NOMO) theory \cite{tachikawa_1998, nakai_2002, nakai_2003, nakai_2007,nomo-new1,nomo-new2,nomo-new3} (other authors denote similar theories with different acronyms, e.g. ENMO \cite{bochevarov_2004} or NEO \cite{webb_2002}) is an extension of MO theory to the non-BOA problem, which employs the idea of a nuclear orbital (NO), as a one-particle orbital of a nucleus. Since electrons and nuclei are treated on equal footing in NOMO framework, it goes beyond the Born-Oppenheimer and adiabatic approximations \cite{born_oppenheimer}. 

In the NOMO method, Gaussian-type functions are adopted for electronic as well as nuclear basis functions, which leads to difficulties in gauging the total-energy accuracy because of the poor description of translational and rotational motions. For this reason, translation- and rotation-free (TRF) approach has been developed by eliminating the contribution of translation and rotation from the total Hamiltonian \cite{nomo-new1,nomo-new2,nomo-new3,nakai_2007}. The TRF Hamiltonian has the form

\begin{eqnarray}
	\hat{H}_{\rm TRF} & = & \hat{H} - \hat{T}_{\rm T} - \hat{T}_{\rm R} \nonumber \\
	 & = & \hat{T}_{\rm TRF}^{\rm n} + \hat{T}_{\rm TRF}^{\rm e} + \hat{V}_{\rm TRF}^{\rm nn} + \hat{V}_{\rm TRF}^{\rm ne} + \hat{V}_{\rm TRF}^{\rm ee}.
	\label{TRF.0}
\end{eqnarray}

\noindent
The total translational Hamiltonian $\hat{T}_{\rm T}$ is simply subtracted, but since molecular rotations and vibrations are coupled, the contribution of rotation cannot be entirely eliminated. 
The only viable way how to subtract rotation is by Taylor expansion of the rotational Hamiltonian $\hat{T}_R$ with respect to $\Delta r_{\mu}$ (as defined in \cite{nomo-new3}). In our numerical study (see Section \ref{section_application}), this was done just to zeroth order.

The Hamiltonian $\hat{H}_{\rm TRF}$ contains one-particle (nucleus: $\hat{T}_{\rm TRF}^{\rm n}$, electron: $\hat{T}_{\rm TRF}^{\rm e}$) and two-particle (nucleus-nucleus: $\hat{V}_{\rm TRF}^{\rm nn}$, nucleus-electron: $\hat{V}_{\rm TRF}^{\rm ne}$, electron-electron: $\hat{V}_{\rm TRF}^{\rm ee}$) terms and has the same second-quantized form as in Eq. \ref{ham_sec_quant}. If, for simplicity, we consider now only one kind of nucleus and use big subscripts $\{P, Q, \ldots\}$ for NOs and small $\{p, q, \ldots\}$ for MOs, we can write

\begin{eqnarray}
 \hat{H}_{\rm TRF} & = & \sum_{pq} h_{pq}^{\rm ee} \hat{a}_{p}^{\dagger} \hat{a}_{q} + \sum_{PQ} h_{PQ}^{\rm nn} \hat{a}_{P}^{\dagger} \hat{a}_{Q} \nonumber \\
  & + & \frac{1}{2} \sum_{pqrs} V_{pqrs}^{\rm ee} \hat{a}_{p}^{\dagger} \hat{a}_{q}^{\dagger} \hat{a}_{s} \hat{a}_{r} + \frac{1}{2} \sum_{PQRS} V_{PQRS}^{\rm nn} \hat{a}_{P}^{\dagger} \hat{a}_{Q}^{\dagger} \hat{a}_{S} \hat{a}_{R} \nonumber \\
  & + & \sum_{pQrS} V_{pQrS}^{\rm en} \hat{a}_{p}^{\dagger} \hat{a}_{Q}^{\dagger} \hat{a}_{S} \hat{a}_{r}.
  \label{ham_sec_quant_trf}
\end{eqnarray}

In principle, the NOMO/FCI theory for a complete configuration space is an exact theory. In practice however, due to the exponential scaling of classical FCI, some approximation has to be adopted. Different kinds of NOMO post-Hartree-Fock methods analogous to those from the conventional Born-Oppenheimer electronic structure theory has been developed, to name a few e.g. NOMO/MP2 \cite{nakai_2003, nomo-new2}, NOMO/CI \cite{nakai_2001}, or NOMO/CC \cite{nakai_2003}. 
One of the drawbacks of the NOMO theory is a slow convergence of the n-e correlation effect with respect to CI/CC expansion \cite{hoshino_2011,nomo-new4,nomo-new5}. 
As we are dealing with the NOMO/FCI theory here, which is, as already mentioned, an exact theory, we avoid this problem. For more details about the NOMO methodology and the discussion of its pros and cons, we refer the reader to the original literature \cite{nakai_2007, hoshino_2011}.

\section{Quantum NOMO/FCI algorithm}
\label{section_nomo_qfci}

In this section we elaborate how the qFCI algorithm is employed for the NOMO Hamiltonian (\ref{ham_sec_quant_trf}).
Despite restricting ourselves to distinguishable nuclei in the proof-of-principle numerical simulations, the presentation in this section is intentionally kept as general as possible, i.e. considering several types of fermionic and/or bosonic nuclei.

The NOMO Hamiltonian (\ref{ham_sec_quant_trf}) can be cast to the general form (\ref{ham_sec_quant}), if the indices $p,q,r,s$ are allowed to run over nuclei spin orbitals as well, i.e.  $h_{pq}$ and $V_{pqrs}$ stands for one and two-particle integrals over molecular and/or nuclear spin orbitals.
For this we define the following ordering of the electronic and nuclear spin orbitals.
Let us consider a general system consisting of $K$ kinds of different nuclei, where we have  $N_0$ molecular spin orbitals for $n_0$ electrons, $N_1$ nuclear spin orbitals for $n_1$ nuclei of the 1st kind, etc., up to $N_K$ nuclear spin orbitals for $n_K$ nuclei of the $K$-th kind.
Say that $K_{\rm{ferm}}$ kinds correspond to fermionic nuclei and the rest to bosonic ones.

Therefore $p,q,r,s \in \{1, 2, ..., N_T\}$, where $N_T$ is the total number of spin orbitals

\begin{equation}
	N_T = \sum_{k=0}^{K} N_k.
	\label{sh1}
\end{equation}

\noindent
The operator $\hat{a}_{p}^{\dagger}$ in (\ref{ham_sec_quant}) denotes the creation operator for $p$-th spin orbital [creating either electron (for $p \leq N_0$) or nucleus (for $p > N_0$)] and $\hat{a}_q$  is the annihilation operator. 

\noindent
The $k$-th kind nuclear spin orbital indices belong to the set $S_k$

\begin{equation}
	S_k = \{ I_k, I_k + 1, \ldots, I_k + N_k - 1 \},
\end{equation}

\noindent
where $I_k$ is the $k$-th nuclei starting index

\begin{equation}
	I_k = \sum_{i = 0}^{k - 1} N_i + 1.
\end{equation}

All creation and annihilation operators commute with both creation and annihilation operators for different particles 

\begin{eqnarray}
	[\hat{a}_p^{\dagger},\hat{a}_q^{\dagger}] = [\hat{a}_p^{\dagger},\hat{a}_q] = [\hat{a}_p,\hat{a}_q] = 0, \label{different_commute} \\
	\mathrm{if} \quad p \in S_k, q \in S_l \quad \mathrm{and} \quad  k \ne l. \nonumber
\end{eqnarray}

\noindent
If particles of $k$-th kind are fermionic, the usual anticommutation relation holds

\begin{eqnarray}
	\{\hat{a}_p^{\dagger},\hat{a}_q^{\dagger}\} = \{\hat{a}_p,\hat{a}_q\} = 0, \quad \{\hat{a}_p^{\dagger},\hat{a}_q\} = \delta_{pq}, \label{sh4} \\
	\mathrm{if} \quad p,q \in S_k \quad \text{and fermions,} \nonumber
\end{eqnarray}

\noindent
while bosonic nuclei must obey the commutation relations

\begin{eqnarray}
	[\hat{a}_p^{\dagger},\hat{a}_q^{\dagger}] = [\hat{a}_p,\hat{a}_q] = 0, \quad [\hat{a}_p^{\dagger},\hat{a}_q] = \delta_{pq}, \label{sh5} \\
	\mathrm{if} \quad p,q \in S_k \quad \text{and bosons.} \nonumber
\end{eqnarray}

\noindent
Distinguishable identical nuclei can be considered as 
%$N$ particles 
%representing $N$ 
separate classes of different nuclei, each class consisting of a single nucleus, so that
the relations (\ref{different_commute}) apply.

The mapping between (anti)symmetrized products of spin orbitals of our general system 
and states of a quantum register can be constructed in the following way.
Let us order the nuclei classes so that fermionic nuclei ($k \leq K_{\rm{ferm}}$) precedes the bosonic ones ($k > K_{\rm{ferm}}$) and discuss the mapping for different particle types separately.

\begin{comment}
Let $W_n$ be the number of qubits needed to store the occupation number of bosonic spin orbital for particles of $n$-th kind, cf. (\ref{sh6}).
\begin{equation}
U_{n+1}\verb| |\geq\verb| |W_n\verb| |\geq\verb| |W_{n,min}\verb| = |\lceil\text{log}_2(U_n+1)\rceil. 
\label{sh6}
\end{equation}

Now, state with occupation numbers $f(n)$ is represented by computational basis state of quantum register (with $Q_{max} + W_{Q+1}N_{Q+1} + W_{Q+2}N_{Q+2} + ... + W_K N_K$ qubits divided to $Q_{max}$ subregisters of one qubit, $N_{Q+1}$ subregisters of $W_{Q+1}$ qubits, $N_{Q+2}$ subregisters of $W_{Q+2}$ qubits, ..., $N_K$ subregisters of $W_K$ qubits)
 which correspond to the direct product

\begin{equation}
|\psi\rangle\verb| = |(\otimes_{k=1}^{Q_{max}}|f(k)\rangle_1)\bigotimes(\otimes_{m=1}^{K-Q}\otimes_{k=(Q+m)_{min}+1}^{(Q+m)_{max}}|f(k)\rangle_{W_{Q+m}}).
\label{sh7}
\end{equation}

Let us denote  $\hat{A}$  a general creation or annihilation operator and $\phi(\hat{A})$ its image  acting on the qubit register and discuss the mapping for different particle types separately.
\end{comment}

\subsubsection{Fermions}

For fermions of $k$-th kind ($p \in S_k$), the standard Jordan-Wigner mapping (\ref{jordan-wigner}) can be used, which
in our index convention can be expressed as

\begin{eqnarray}
	\hat{a}_{p}^{\dagger} & = & \Bigg( \bigotimes_{\substack{i \in S_k \land \\ i < p}} \sigma_{z}^{i} \Bigg) \otimes \sigma_{-}^{p}, \label{ferm1} \\
	 \hat{a}_{p} & = & \Bigg( \bigotimes_{\substack{i \in S_k \land \\ i < p}} \sigma_{z}^{i} \Bigg) \otimes \sigma_{+}^{p}. \label{ferm2}
\end{eqnarray}

%\begin{equation}
%	\phi(\hat{a}_i^{\dagger})\verb| = |\sigma_{z,n_{min}+1}\sigma_{z,n_{min}+2}\cdots\sigma_{z,i-1}\sigma_{-,i}
%	\label{sh7b}
%\end{equation}
%and
%\begin{equation}
%\phi(\hat{a}_i)\verb| = |\sigma_{z,n_{min}+1}\sigma_{z,n_{min}+2}\cdots\sigma_{z,i-1}\sigma_{+,i},
%\label{sh7c}
%\end{equation}

\noindent
When individual $\exp(i\hat{h}_X\tau/N)$ (see Eq. \ref{trotter}) terms are implemented, this approach 
%increases the $\mathcal{O}(N_k^4)$ Hamiltonian complexity to $\mathcal{O}(N_k^5)$. 
leads to the computational cost $\mathcal{O}(N_k^5)$.
Alternatively, the more economic Bravyi-Kitaev transformation \cite{bravyi_2002, seeley_2012} [$\mathcal{O}(N_k^4 \log{N_k})$] can be employed.

\subsubsection{Bosons}
In contrast to fermions, more than one qubit is needed to store an occupation number of a bosonic spin orbital. When considering bosonic particles of $k$-th kind, the occupation number $f(p)$ of the spin orbital $p$ can acquire values from $0$ to $n_k$. 
The so called direct boson mapping \cite{somma_2002} uses $n_k + 1$ qubits to store this occupation number in the following way
 
\begin{equation}
	\ket{f(p)} = \Bigg( \bigotimes_{i=0}^{f(p)-1} \ket{0}_i \Bigg) \otimes \ket{1}_{f(p)} \otimes \Bigg( \bigotimes_{i=f(p)+1}^{n_k} \ket{0}_i \Bigg).
	\label{sh7.3}
\end{equation}

\noindent
For clarity, we take into account only the part of a quantum register that corresponds to $p$-th spin orbital (in bosonic case one creation or annihilation operator does not act on qubits corresponding to different spin orbitals). 

From (\ref{sh5}) follows the action of creation and annihilation operators 

\begin{eqnarray}
	\hat{a}_p^{\dagger} |f(p)\rangle & = & \sqrt{f(p)+1} |f(p)+1\rangle, \label{sh7.5} \\
	\hat{a}_p |f(p)\rangle & = & \sqrt{f(p)} |f(p)-1\rangle, \label{sh7.6} \\
	%\hat{a}_p^{\dagger} |N_k\rangle & = & 0, \label{sh7.7b} \\
	\hat{a}_p |0\rangle & = & 0. \label{sh7.7}
\end{eqnarray}

\noindent
As the maximum occupation $f(p)$ is the number of particles $n_k$, one more condition has to be introduced

\begin{equation}
	\hat{a}_p^{\dagger} |n_k\rangle = 0.
	\label{sh7.11}
\end{equation}

In analogy to Eqs. (\ref{ferm1}) and (\ref{ferm2}), Somma \textit{et al.} \cite{somma_2002} proposed the direct boson mapping of the form

\begin{eqnarray}
	\hat{a}_p^{\dagger} & = & \sum_{j=0}^{n_k-1} \sqrt{j+1} \Big( \sigma_{+}^{j} \otimes \sigma_{-}^{j+1} \Big), \label{sh7.12} \\
	\hat{a}_p           & = & \sum_{j=0}^{n_k-1} \sqrt{j+1} \Big( \sigma_{-}^{j} \otimes \sigma_{+}^{j+1} \Big). \label{sh7.12b}
\end{eqnarray}

\noindent
Following their complexity analysis, we can express the overall computational cost of a single Trotter step of the general system as

%in \cite{somma_2002} of the $\exp(i V_{pqrs}\tau\{\hat{a}_p^{\dagger}\hat{a}_q^{\dagger}\hat{a}_s\hat{a}_r+h.c.\})$  part of the Trotter-Suzuki expansion of molecular Hamiltonian has been performed, showing that the computational cost 
%of implementing the evolution operator in the direct boson mapping is

\begin{gather}
 \mathcal{O} \Bigg(\sum_{0 \leq k \leq l \leq K_{\mathrm{ferm}}} N_k^2 N_l^2 \log(N_k N_l)\Bigg) + \nonumber \\
	 + \mathcal{O} \Bigg(\sum_{\substack{K_{\mathrm{ferm}} < k \leq l \leq K}} n_k^2 n_l^2 N_k N_l^2\Bigg) + \nonumber \\
	+ \mathcal{O}\Bigg(\sum_{\substack{0 \leq k \leq K_{\mathrm{ferm}} \\ K_{\mathrm{ferm}} < l \leq K}} n_l^2 \log(N_k) N_l N_k^2 \Bigg),
	\label{sh7.13}
\end{gather}

\noindent
where individual terms correspond to contributions from fermion-fermion, boson-boson and fermion-boson interactions, respectively, and $N_k \leq N_l$ for each $K_{\mathrm{ferm}}+1 \leq k < l$ is supposed for simplicity. The Bravyi-Kitaev transformation is considered for fermions in (\ref{sh7.13}), as well as in the following formulae (\ref{sh15}) and (\ref{sh15n}).

In contrast to the direct boson mapping, we propose the compact boson mapping which uses only 
$\lceil\text{log}_2(n_k+1)\rceil$  qubits to represent the 
binary expansion of $f(p)$, as shown below (the least significant bit is written leftmost)

\begin{eqnarray}
	|f(p)\!=\!0\rangle & \quad \mapsto \quad & |0\rangle|0\rangle|0\rangle...|0\rangle, \nonumber \\
	|f(p)\!=\!1\rangle & \quad \mapsto \quad & |1\rangle|0\rangle|0\rangle...|0\rangle, \nonumber \\
	|f(p)\!=\!2\rangle & \quad \mapsto \quad & |0\rangle|1\rangle|0\rangle...|0\rangle, \nonumber \\
	|f(p)\!=\!3\rangle & \quad \mapsto \quad & |1\rangle|1\rangle|0\rangle...|0\rangle, \text{etc.}
	%|f(i)=k\rangle & \mapsto & |k\rangle_{q,i}=|f(i)=k\rangle_{W_{n,min}},\verb|   |1\leq k \leq U_n.
\end{eqnarray}

The representation of creation and annihilation operators for $p$-th spin orbital is defined implicitly by 
relations (\ref{sh7.5})-(\ref{sh7.11}) and we can express them semi-formally by the formulae

\begin{eqnarray}
        \hat{a}_p^{\dagger} & \quad = \quad & \sum_{j=0}^{n_k-1} \sqrt{j+1} \; |f(p)\!=\!j+1\rangle \; \langle f(p)\!=\!j| , \label{sh7.12aa} \\
        \hat{a}_p           & \quad = \quad & \sum_{j=0}^{n_k-1} \sqrt{j+1} \; |f(p)\!=\!j\rangle \; \langle f(p)\!=\!j+1|. \label{sh7.12baa}
\end{eqnarray}

\noindent

Rather than using formulae (\ref{sh7.12aa}) and (\ref{sh7.12baa}) explicitly, circuit representation of appropriate combinations of the bosonic creation and annihilation operators occuring in (\ref{ham_sec_quant_trf}) 
were investigated.
In the Appendix we show that this approach leads to the overall computational cost of a single Trotter step

\begin{gather}
 \mathcal{O} \Bigg( \sum_{0 \leq k \leq l \leq K_{\mathrm{ferm}}} N_k^2 N_l^2 \log(N_k N_l)\Bigg) + \nonumber \\
    \mathcal{O} \Bigg( \sum_{\substack{K_{\mathrm{ferm}} < k \leq l \leq K}} N_{\mathrm{g}}(n_k,n_l) N_k N_l^2 \Bigg) + \nonumber \\
 + \mathcal{O} \Bigg( \sum_{\substack{0 \leq k \leq K_{\mathrm{ferm}} \\ K_{\mathrm{ferm}} < l \leq K}} N_{\mathrm{g}}(1,n_l) \log(N_k) N_l N_k^2 \Bigg),
	\label{sh15}  % BKT used for fermions
\end{gather}

\noindent
where

\begin{gather}
N_{\mathrm{g}}(n,m)= -\frac {1}{15} y + \frac {1}{3} x y (x+y) + \nonumber \\
+ \frac {2}{3} x^2 y^3 - \frac {1}{3} x y^4 + \frac {1}{15} y^5,
\label{ngat}
\end{gather}

\noindent
$x = \max(n,m) + 1$, $y = \min(n,m) + 1$. The derivation of (\ref{sh15}) (See Appendix) took into account that bosonic terms with non-overlaping sets of indices can be processed simultaneously. 

In the simplified case where $n_k = n$ for each $k > K_{\mathrm{ferm}}$
$N_{\mathrm{g}}(n,n)=\mathcal{O} (n^5)$ and $N_{\mathrm{g}}(1,n)=\mathcal{O} (n^2)$ and the single Trotter step overall computational cost is

\begin{gather}
 \mathcal{O} \Bigg( \sum_{0 \leq k \leq l \leq K_{\mathrm{ferm}}} N_k^2 N_l^2 \log(N_k N_l) \Bigg) + \nonumber \\
        + \mathcal{O} \Bigg( \sum_{\substack{K_{\mathrm{ferm}} < k \leq l \leq K}} n^5 N_k N_l^2 \Bigg) + \nonumber \\
 + \mathcal{O} \Bigg( \sum_{\substack{0 \leq k \leq K_{\mathrm{ferm}} \\ K_{\mathrm{ferm}} < l \leq K}} n^2 \log(N_k) N_l N_k^2 \Bigg).
        \label{sh15n}  % BKT used for fermions
\end{gather}

%\noindent
%where $M_k = N_k n_k$ is the number of ``particle-states''.
% - number of spin orbitals needed for a fermionic particle ``of the same quality'' due to the Pauli principle.

\subsubsection{Distinguishable particles}

Distinguishable particles can be distributed to separate classes each with $n_k = 1$ and occupation numbers $f(p) \in \{0;1\}$ with $p \in S_k$ can be stored using one qubit per spin orbital for each of the particles.
%If the particles of $k$-th kind are distinguishable, we can suppose that $n_k = 1$ and thus use $N_k$ qubits to store the occupation numbers $f(p) \in \{0;1\}$ with $p \in S_k$. 
The creation and annihilation operators can then be represented as

%For n-th kind of particles being distinguishable it is usual to suppose that $U_n = 1$ therefore we can use the $N_n$ qubits to store the 
%occupation numbers $f(i) \in \{0;1\}$ of spin orbitals with indices $i \in \{n_{min} + 1, n_{min} + 2, ..., n_{max}\}$, the corresponding operator algebra 
%mapping will be (compare with (\ref{sh7b}) and (\ref{sh7c}))

\begin{eqnarray}
	\hat{a}_p^{\dagger} & = & \sigma_{-}^{p}, \label{sh16b} \\
	\hat{a}_p & = & \sigma_{+}^{p} \label{sh16c}
\end{eqnarray}

\noindent
and we can use either (\ref{sh7.13}) or (\ref{sh15n}), which are in fact equivalent in this case as $n_k = 1$.
%, as $M_k = N_k$ if $n_k = 1$. 
% for case where particles for $n > Q$ particles are either \textit{bosonic} or \textit{distinguishable} (we can chose some $S \geq Q$ so that for  $S > n > Q$ particles are bosonic and for $n > S$ distinguishable). 
%Since for $n_k = 1$ number of ``particle-states'' 
% $M_k$ equals to $N_k$, the corresponding terms ($n > S$) are the same for (\ref{sh7.13}) and (\ref{sh15}) (differing just by unimportant prefactor $\alpha$  or $\tilde{\alpha}$ ). \\

\vskip 0.2 cm

The equations (\ref{sh7.13}) and (\ref{sh15n}) demonstrate that the qFCI algorithm can be \textit{efficient} also for systems with bosonic and/or distinguishable particles.
 In case of the compact boson mapping, the smaller number of qubits needed to represent a state of a molecule (as compared with the direct boson mapping) 
is paid by worse computational cost in terms of the number of two-qubit gates needed for a computation. For the first, few-qubit quantum computers,
 the compact boson mapping should be important, however for larger scale quantum computers, the direct boson mapping is of much greater importance.

We would like to note that presented mappings are not restricted to (I)PEA algorithm only. They, in fact, can be also used in connection with other methods that were developed to reduce qubit and coherence time requirements \cite{lanyon_2010, biamonte_2011, li_2011, peruzzo_2013}, i.e. to adapt the procedure for a present-day or near-future quantum technology. 

\section{Application to H$_2$ and HT molecules}
\label{section_application}

%\begin{itemize}
%  \color{red}
%  \item comment on spurious states and refer to more detailed discussion in \cite{bochevarov_2004}
%  \item exchange interaction between nuclei is negligibly small and they thus can be considered as independent particles
%\end{itemize}

\subsection{Computational details}

%Molecular basis integrals $h_{ij}$ and $V_{ijkl}$ (\ref{secqham}) were obtained from TRF-NOMO/Restricted
%Hartree-Fock calculation done with cc-pVTZ \cite{basis_dunning_1989} ABS consisting three s-functions
%(from which one was contracted from three primitive gaussians and the others were not contracted)
%two p-functions and one d-function and NBF consisting of one s, one p and one d non-contracted gaussians with five even-tempered exponents
%from 9.081045 to 908.104502 (50 NBF functions for each hydrogen atom).
%Both ABS and NBF were centred in nuclear equilibrium positions $R$ = 0.750746 \AA\verb| | apart each other. The nuclei were for simplicity treated as distinguishable.  % 2015-03-16 18:22
%We were interested in states with $M_S = 0$ (zero total $z$-component of electron spin) and therefore the two molecular spin orbitals are of different spin-part leading to electron singlet (upper sign) or electron triplet (lower sign) combinations  % 2015-03-16 18:22

For the proof-of-principle classical simulations, we have chosen the simplest molecular examples, namely the two isotopomers of the hydrogen molecule (H$_2$, HT). For MOs, we employed the cc-pVTZ basis set, while NOs were expanded in a basis of one $s$, one $p$, and one $d$ non-contracted Gaussians centred on each hydrogen atom with five even-tempered exponents from 9.081045 to 908.104502 for H and from 27.18608 to 2718.608 for T isotope (total 50 nuclear basis functions for each hydrogen atom). 
The internuclear distance was fixed to $R$ = 0.750746 \AA. 

The nuclei were for simplicity treated as distinguishable which is the usual procedure that can be justified by the fact that exchange interaction between nuclei is negligibly small. 
%We were interested in states with $M_S = 0$ (zero $z$-component of total electron spin) and therefore the two molecular spin orbitals are of different spin part leading to electron singlet (upper sign) or electron triplet (lower sign) combinations.

%The simulation of the quantum algorithm on classical computer employed a compact mapping of Hilbert spaces and $\hat{U}$
%was therefore represented as a single $l$-qubit gate
%($l = \ceil*{log_{2}(d)}$, where $d$ is dimension of FCI hamiltonian matrix for hydrogen molecule,
%$d = N^2 \cdot M^2$. $N$ and $M$ being the number of molecular and nuclear orbitals used in the FCI hamiltonian matrix
%construction).
%Using our C++ implementation, we simulated $m = 17$ iterations of both IPEA A and IPEA B with input parameters
%$E_{\rm min} = -1.20\verb| | a.u.$ and $E_{\rm max} = -1.00\verb| | a.u.$.
%The largest simulation for $M = 23$ and $N = 9$ took us 3 weeks CPU-time on Intel Xeon 3GHz.

We worked solely with a compact mapping from subspace of wave functions with zero $z$-component of total electron and nuclear spins and the exponential of a Hamiltonian was simulated as an $n$-qubit gate (similarly as in Refs. \cite{aspuru-guzik_2005, wang_2008, veis_2010, veis_2012}). We simulated $m = 17$ iterations of both IPEA \textbf{A} and \textbf{B} with input parameters (see Ref. \cite{veis_2010})
$E_{\rm min} = -1.20$ a.u. and $E_{\rm max} = -1.00$ a.u.

The ground state of both H$_{2}$ and HT is dominated by $1\sigma_g^2 1\sigma_{n1} 1\sigma_{n2}$ configuration, while
the excited state $\nu = 1, J = 0$ was identified as a state dominated by $1\sigma_g^2 2\sigma_{n1} 2\sigma_{n2}$ configuration.
Between all states in $\pm 50\%$ interval around the experimental transition energy, only this state is non-degenerate and symmetric with respect to
the exchange of the nuclear coordinates
(and therefore has even rotation number $J$ and is a nuclear singlet).

The initial guesses of both states were single determinants (ground state: $1\sigma_g^2 1\sigma_{n1} 1\sigma_{n2}$, excited state: $1\sigma_g^2 2\sigma_{n1} 2\sigma_{n2}$).

However, we must note that for larger polyatomic molecules the identification of several rotationless vibrational excited states and the choice of
sufficiently accurate initial
guesses of eigenvectors for IPEA might be much more complicated than it was for the hydrogen molecule.

\subsection{Results}

In Table \ref{Tabc1}, we show energies and IPEA (\textbf{A}) success probabilities for the ground and excited states of H$_2$, while Table \ref{Tabc2}
 presents
  minimal number of repetitions of IPEA \textbf{A} and IPEA \textbf{B} needed to achieve a given success probability (0.99, 0.999 999).
Tables \ref{Tabc3} and \ref{Tabc4} give corresponding information for the HT molecule.

\begin{table*}[!ht]
  \begin{tabular}{c | c |  l | l | c | c | c | r | c | c | c | c}
  \hline
	\hline
 $N$ & $M$ & MOs & NOs & $ E_{\rm gs} $ & $ E_{\rm es} $ & $\Delta E$ & $\delta\omega$ & $ S_{\rm gs} $ & $ S_{\rm es} $ & $ p_{\rm A,gs} $ & $ p_{\rm A,es} $ \\
 & & & & a.u. & a.u. & cm${}^{-1}$ & & & & \\
\hline
1 & 4 & $\sigma_{\rm g}$ & $ 2\sigma\pi $ & -1.104049 & -1.080972 & 5064.8 & 21.7 & 0.9985 & 0.9164 & 0.8217 & 0.8743 \\
2 & 6 & $\sigma_{\rm g}\sigma_{\rm u}$ & $ 2\sigma\pi\delta $ & -1.105483 & -1.082857 & 4965.7 & 19.3 & 0.9973 & 0.6737 & 0.9645 & 0.5760 \\
3 & 8 & $2\sigma_{\rm g}\sigma_{\rm u}$ & $ 2\sigma2\pi\delta $ & -1.108131 & -1.085068 & 5061.7 & 21.6 & 0.9880 & 0.7784 & 0.8073 & 0.7780 \\
4 & 10 & $2\sigma_{\rm g}2\sigma_{\rm u}$ & $4\sigma2\pi\delta $ & -1.119317 & -1.096120 & 5091.1 & 22.3 & 0.9759 & 0.8344 & 0.8769 & 0.6779 \\
6 & 10 & $2\sigma_{\rm g}2\sigma_{\rm u}\pi_{\rm u}$ &  & -1.127425 & -1.107187 & 4441.8 & 6.7 & 0.9583 & 0.9385 & 0.9419 & 0.8454 \\
6 & 15 &  & $5\sigma3\pi2\delta $ & -1.127443 & -1.107373 & 4404.8 & 5.9 & 0.9575 & 0.9282 & 0.9329 & 0.9102 \\
6 & 23 &  & $7\sigma5\pi3\delta $ & -1.127612 & -1.108051 & 4293.3 & 3.2 & 0.9571 & 0.9135 & 0.9565 & 0.9095 \\
7 & 18 & $3\sigma_{\rm g}2\sigma_{\rm u}\pi_{\rm u}$ & $6\sigma3\pi3\delta $ & -1.129852 & -1.109358 & 4497.9 & 8.1 & 0.9404 & 0.9041 & 0.8113 & 0.8056 \\
7 & 23 &  & $7\sigma5\pi3\delta $ & -1.129917 & -1.110242 & 4318.0 & 3.8 & 0.9464 & 0.8900 & 0.9059 & 0.7301 \\
9 & 10 & $3\sigma_{\rm g}2\sigma_{\rm u}\pi_{\rm u}\pi_{\rm g}$ & $4\sigma2\pi\delta$  & -1.130100 & -1.109888 & 4436.0 & 6.6 & 0.9419 & 0.9053 & 0.7718 & 0.7387 \\
9 & 18 &  & $6\sigma3\pi3\delta$ & -1.130285 & -1.110317 & 4382.5 & 5.3 & 0.9334 & 0.8779 & 0.7971 & 0.8735 \\
9 & 20 &  & $6\sigma4\pi3\delta$ & -1.130293 & -1.110419 & 4361.9 & 4.8 & 0.9342 & 0.8880 & 0.9316 & 0.8455 \\
9 & 23 &  & $7\sigma5\pi3\delta$ & -1.130346 & -1.111238 & 4193.7 & 0.8 & 0.9398 & 0.8600 & 0.8376 & 0.7394 \\
  &    & Theory non-rel.\footnote{Accurate non-relativistic beyond-BOA theoretical values\cite{stanke_2008,bubin_2009,pachucki_2009}.} 
  &   & -1.164025 & -1.145065 & 4161.2 &  &         &        &        &        \\ 
  &    & Theory rel.\footnote{Accurate relativistic beyond-BOA theoretical values based on dissociation energy $D_{0,\rm{theo}}$ = 36118.0696(11) cm${}^{-1}$ = 0.1645660 a.u. \cite{komasa_2011},
rotationless 0-1 vibrational transition energy $\Delta E_{\rm{theo}}$ = 4161.1661(9) cm${}^{-1}$ \cite{komasa_2011}, 
1${}^{2}$S$_{1/2}$ dirac energy for free hydrogen
atoms, $E(H)$ = -0.4997345 a.u., corrected by Lamb shift $E_{\rm{LS,theo}}(H)$ = 8172.802(40) MHz = 0.0000012 a.u. taken from 
\cite{weitz_1995}. For the ground state $E_{\rm{gs,theo}} = 2(E(H)+E_{\rm{LS,theo}}(H))-D_{0,\rm{theo}}$, and for the excited state $E_{\rm{es,theo}} = E_{\rm{gs,theo}} + \Delta E_{\rm{theo}}$.} &   & -1.164033 & -1.145073  & 4161.2 &  &         &        &        &        \\
  &    & Experiment\footnote{Experimental values based on the
dissociation energy $D_{0,\rm{exp}}$ = 36118.06962(37) cm${}^{-1}$ \cite{liu_2009}, rotationless 0-1 vibrational transition energy $\Delta E_{\rm{exp}}$ = 4161.16632(18) cm${}^{-1}$ \cite{dickenson_2013}
and 1${}^{2}$S$_{1/2}$ hydrogen atom Lamb shift $E_{\rm{LS,exp}}$ = 8172.874(60) MHz \cite{weitz_1995}. For the ground state 
$E_{\rm{gs,exp}} = 2(E(H)+E_{\rm{LS,exp}}(H))-D_{0,\rm{exp}}$, and for the excited state $E_{\rm{es,exp}} = E_{\rm{gs,exp}} + \Delta E_{\rm{exp}}$. Values are without hyperfine splitting.} &   & -1.164033 & -1.145073  & 4161.2 &  &         &        &        &        \\
  \hline
	\hline
  \end{tabular}
  \caption{Ground ($E_{\rm gs}$) and excited state ($E_{\rm es}$, $\nu=1$, $J=0$) NOMO-TRF/FCI energies and IPEA (\textbf{A}) success probabilities ($p_{\rm A,gs}$, $p_{\rm A,es}$) for the H$_{2}$ molecule 
in the basis consisting of $N$ molecular orbitals and $M$ nuclear orbitals. $\Delta E$ denotes transitional energy, $\delta\omega = 100 (\Delta E-\Delta E_{\rm exp})/\Delta E_{\rm exp}$,
  $S_{\rm gs}$ and $S_{\rm es}$ denote square of absolute value of the overlaps between NOMO-TRF/FCI eigenvectors and their initial guesses.
}
  \label{Tabc1}
\end{table*}

\begin{table*}
  \begin{tabular}{r | r |  l | l | r | r | r | r | r | r | r | r}
  \hline
	\hline
 $N$ & $M$ & MOs & NOs & $ N_{\rm{B},\rm{gs},2} $ & $ N_{\rm{A},\rm{gs},2} $ & $N_{\rm{B},\rm{gs},6}$ & $N_{\rm{A},\rm{gs},6}$ & $ N_{\rm{B},\rm{es},2} $ & $ N_{\rm{A},\rm{es},2} $ & $N_{\rm{B},\rm{es},6}$ & $N_{\rm{A},\rm{es},6}$ \\
\hline
1 & 4 & $\sigma_{\rm g}$ & $ 2\sigma\pi $ &7&7&37&29&7&7&21&23 \\
2 & 6 & $\sigma_{\rm g}\sigma_{\rm u}$ & $ 2\sigma\pi\delta $    &3&3&11&7&39&63&131&253 \\
3 & 8 & $2\sigma_{\rm g}\sigma_{\rm u}$ & $ 2\sigma2\pi\delta $   &9&7&37&29&13&9&45&35 \\
4 & 10 & $2\sigma_{\rm g}2\sigma_{\rm u}$ & $4\sigma2\pi\delta $    &5&5&25&15&11&13&41&39 \\
6 & 10 & $2\sigma_{\rm g}2\sigma_{\rm u}\pi_{\rm u}$ &     &5&5&13&11&7&5&29&17 \\
6 & 15 &  & $5\sigma3\pi2\delta $    &5&5&13&13&5&5&17&13 \\
6 & 23 &  & $7\sigma5\pi3\delta $   &5&3&13&11&7&5&19&15 \\
7 & 18 & $3\sigma_{\rm g}2\sigma_{\rm u}\pi_{\rm u}$ & $6\sigma3\pi3\delta $   &7&7&37&23&9&7&37&21 \\
7 & 23 &  & $7\sigma5\pi3\delta $   &5&5&15&15&11&11&43&35 \\
9 & 10 & $3\sigma_{\rm g}2\sigma_{\rm u}\pi_{\rm u}\pi_{\rm g}$ & $4\sigma2\pi\delta$   &9&9&39&31&11&9&55&33 \\
9 & 18 &  & $6\sigma3\pi3\delta$ &9&9&47&27&7&5&23&13 \\
9 & 20 & & $6\sigma4\pi3\delta$ &5&5&17&11&7&5&23&15 \\
9 & 23 & & $7\sigma5\pi3\delta$  &7&7&27&19&11&9&43&27 \\
\hline
\hline
  \end{tabular}
  \caption{
 Minimal number of repetions for IPEA \textbf{B} to achieve success
probability at least $p$ = 0.99 ($N_{\rm{B},y,2}$) or at least $p$ = 0.999 999 ($N_{\rm{B},y,6}$) and the same quantities for IPEA \textbf{A} ($N_{\rm{A},y,2}$, $N_{\rm{A},y,6}$), where $y \in \{\rm{gs} (\text{ground state}), \rm{es} (\text{excited state})\}$.
All data corresponds to H$_2$ molecule.
 }
  \label{Tabc2}
\end{table*}

\begin{table*}
  \begin{tabular}{r | r |  l | l | c | c | c | r | c | c | c | c}
  \hline
	\hline
 $N$ & $M$ & MOs & NOs & $ E_{\rm gs} $ & $ E_{\rm es} $ & $\Delta E$ & $\delta\omega$ & $ S_{\rm gs} $ & $ S_{\rm es} $ & $ p_{\rm A,gs} $ & $ p_{\rm A,es} $ \\
 & & & & a.u. & a.u. & cm${}^{-1}$ & & & & \\
\hline

1&4&  $\sigma_{\rm g}$ & $ 2\sigma\pi $                                   &-1.109233 &-1.091291 &3937.9 &14.6  &0.9989 &0.9171 &0.9968 &0.8171 \\
2&6&  $\sigma_{\rm g}\sigma_{\rm u}$ & $ 2\sigma\pi\delta $               &-1.11064  &-1.092828 &3909.1 &13.8  &0.9977 &0.8853 &0.9769 &0.8736 \\
3&7 & $2\sigma_{\rm g}\sigma_{\rm u}$ & $ 2\sigma2\pi\delta $             &-1.112491 &-1.094738 &3896.2 &13.4  &0.9946 &0.8903 &0.9784 &0.7993 \\
4&10&  $2\sigma_{\rm g}2\sigma_{\rm u}$ & $4\sigma2\pi\delta $            &-1.124135 &-1.106038 &3971.8 &15.6  &0.9766 &0.8359 &0.9475 &0.782 \\
6&10& $2\sigma_{\rm g}2\sigma_{\rm u}\pi_{\rm u}$ &                       &-1.13208  &-1.116166 &3492.7 &1.7   &0.9579 &0.9294 &0.9393 &0.7986 \\
6&15 & & $5\sigma3\pi2\delta $                                             &-1.132094 &-1.11623  &3481.8 &1.3   &0.9571 &0.9175 &0.9513 &0.834 \\
6&23 & & $7\sigma5\pi3\delta $                                             &-1.132224 &-1.116776 &3390.5 &-1.3  &0.957  &0.9016 &0.7987 &0.8004 \\
7&18& $3\sigma_{\rm g}2\sigma_{\rm u}\pi_{\rm u}$ & $6\sigma3\pi3\delta $ &-1.134359 &-1.118117 &3564.7 &3.8   &0.9369 &0.8881 &0.7712 &0.8433 \\
7&23 & & $7\sigma5\pi3\delta $                                             &-1.134412 &-1.118831 &3419.7 &-0.5  &0.9438 &0.88   &0.9396 &0.8304 \\
9&10& $3\sigma_{\rm g}2\sigma_{\rm u}\pi_{\rm u}\pi_{\rm g}$ & $4\sigma2\pi\delta$ &-1.134605 &-1.118688 &3493.3 &1.7   &0.9394 &0.888  &0.8261 &0.7207 \\
9&18 & & $6\sigma3\pi3\delta$                                              &-1.134755 &-1.118924 &3474.5 &1.1   &0.9297 &0.8344 &0.8579 &0.7891 \\
9&20 & & $6\sigma4\pi3\delta$                                              &-1.134757 &-1.119021 &3453.7 &0.5   &0.9303 &0.8501 &0.7582 &0.7359 \\
9&23 & & $7\sigma5\pi3\delta$                                              &-1.134803 &-1.119656 &3324.4 &-3.2  &0.937  &0.8321 &0.7712 &0.7804 \\
& & Theory non-rel.\footnote{Accurate non-relativistic beoyond-BOA theoretical ground state energy \cite{lloydwilliams_2013}.}
 &           &-1.166002 &          &       & & & & \\
& & Theory rel.\footnote{Accurate relativistic beyond-BOA theoretical values based on dissociation energies \cite{kolos_1968} (Tab. IV in \cite{kolos_1968} with relativistic and radiation correction -0.7 cm${}^{-1}$ suggested in text),
and 1${}^{2}$S$_{1/2}$ dirac energy for free hydrogen and tritium,
energies of 1${}^{2}$S$_{1/2}$ atom electronic states were corrected by Lamb shift taken from \cite{weitz_1995} - 
for tritium $\mu_{\rm{T}}/\mu_{\rm{H}}$ multiplied value was used ($\mu_{\rm{T}}$ and $\mu_{\rm{H}}$ being reduced mass of electron for respective atom systems).}
 &           &-1.166007 &-1.150354 &3435.5 & & & & \\
& & Experiment\footnote{Experimental values - For $\Delta E_{\rm{exp}}$ rotationless 1-0 vibrational transition energy Chuang and Zare \cite{chuang_zare_1987} 
report 0.0082 cm${}^{-1}$ uncertainity while Veirs and Rosenblatt report $\Delta E_{\rm{exp}}$ = (3434.9 $\pm$ 0.1) cm${}^{-1}$.} &           &          &          &3434.8 & & & & \\
\hline
\hline
  \end{tabular}
%  \caption{
%Energy and IPEA A success probability for the HT molecule for NOMO-TRF/Full-CI
%in basis consisting of N molecular orbitals and M nuclear orbitals. NOMO-TRF/CI energy for the ground state $E_{\rm gs}$
%and for the excited $(\nu = 1,J = 0) $ state $E_{\rm es}$, $\Delta E = E_{\rm es} - E_{\rm gs}$ is transitional
% energy,
%$\delta\omega = 100 (\Delta E-\Delta E_{\rm exp})/\Delta E_{\rm exp}$ is the deviation of the computed transitional energy from experimental one in percents of the experimental value.
%$S_{\rm gs}$ = $ |\langle \Phi_{1,1,1,1}|\Psi_{\rm gs}\rangle|^{2} $ is overlap between NOMO-TRF/CI eigenvector
% $|\Psi_{\rm gs}\rangle$. and its inital guess $|\Phi_{1,1,1,1}\rangle$, $S_{\rm es} = |\langle\Phi_{1,1,4,4}|\Psi_{\rm es}\rangle|^2$
%  is overlap between NOMO-TRF/CI eigenvector $|\Psi_{\rm es}\rangle$ and its inital guess $|\Phi_{1,1,4,4}\rangle$,
% $p_{\rm A,gs}$ is success probability for IPEA A in the case of ground state and $p_{\rm A,es}$ is the same quantity for the excited state.}
  \caption{Ground ($E_{\rm gs}$) and excited state ($E_{\rm es}$, $\nu=1$, $J=0$) NOMO-TRF/FCI energies and IPEA (\textbf{A}) success probabilities ($p_{\rm A,gs}$, $p_{\rm A,es}$) for the HT molecule 
in the basis consisting of $N$ molecular orbitals and $M$ nuclear orbitals. $\Delta E$ denotes transitional energy, $\delta\omega = 100 (\Delta E-\Delta E_{\rm exp})/\Delta E_{\rm exp}$,
  $S_{\rm gs}$ and $S_{\rm es}$ denote square of absolute value of the overlaps between NOMO-TRF/FCI eigenvectors and their initial guesses.
}
  \label{Tabc3}
\end{table*}

\begin{table*}
  \begin{tabular}{r | r |  l | l | r | r | r | r | r | r | r | r}
	\hline
  \hline
 $N$ & $M$ & MOs & NOs & $ N_{\rm{B},\rm{gs},2} $ & $ N_{\rm{A},\rm{gs},2} $ & $N_{\rm{B},\rm{gs},6}$ & $N_{\rm{A},\rm{gs},6}$ & $ N_{\rm{B},\rm{es},2} $ & $ N_{\rm{A},\rm{es},2} $ & $N_{\rm{B},\rm{es},6}$ & $N_{\rm{A},\rm{es},6}$ \\
\hline
1 & 4 & $\sigma_{\rm g}$ & $ 2\sigma\pi $  &3&1&5&5&7&7&29&21 \\
2 & 6 & $\sigma_{\rm g}\sigma_{\rm u}$ & $ 2\sigma\pi\delta $ &3&3&9&7&7&7&21&17    \\
3 & 7 & $2\sigma_{\rm g}\sigma_{\rm u}$ & $ 2\sigma2\pi\delta $ &3&3&9&7&9&7&31&21   \\
4 & 10 & $2\sigma_{\rm g}2\sigma_{\rm u}$ & $4\sigma2\pi\delta $ &3&5&13&9&11&7&35&23 \\
6 & 10 & $2\sigma_{\rm g}2\sigma_{\rm u}\pi_{\rm u}$ & &5&5&13&9&9&7&43&21     \\
6 & 15 &  & $5\sigma3\pi2\delta $ &5&5&13&11&7&7&29&19    \\
6 & 23 &  & $7\sigma5\pi3\delta $ &9&7&41&25&9&7&33&19   \\
7 & 18 & $3\sigma_{\rm g}2\sigma_{\rm u}\pi_{\rm u}$ & $6\sigma3\pi3\delta $ &9&9&41&31&7&5&19&15   \\
7 & 23 &  & $7\sigma5\pi3\delta $ &5&5&15&11&7&5&25&15   \\
9 & 10 & $3\sigma_{\rm g}2\sigma_{\rm u}\pi_{\rm u}\pi_{\rm g}$ & $4\sigma2\pi\delta$ &7&7&31&21&11&11&51&35   \\
9 & 18 &  & $6\sigma3\pi3\delta$ &7&5&25&17&11&7&29&21  \\
9 & 20 & & $6\sigma4\pi3\delta$ &11&9&39&33&11&9&41&25  \\
9 & 23 & & $7\sigma5\pi3\delta$ &9&9&41&31&11&7&29&19 \\
\hline
\hline
  \end{tabular}
  \caption{
Minimal number of repetions for IPEA \textbf{B} to achieve success
probability at least $p$ = 0.99 ($N_{\rm{B},y,2}$) or at least $p$ = 0.999 999 ($N_{\rm{B},y,6}$) and the same quantities for IPEA \textbf{A} ($N_{\rm{A},y,2}$, $N_{\rm{A},y,6}$), where $y \in \{\rm{gs} (\text{ground state}), \rm{es} (\text{excited state})\}$.
All data corresponds to HT molecule.
 }
  \label{Tabc4}
\end{table*}

\begin{figure}[!ht]
  \begin{center}
    \includegraphics[width=8cm]{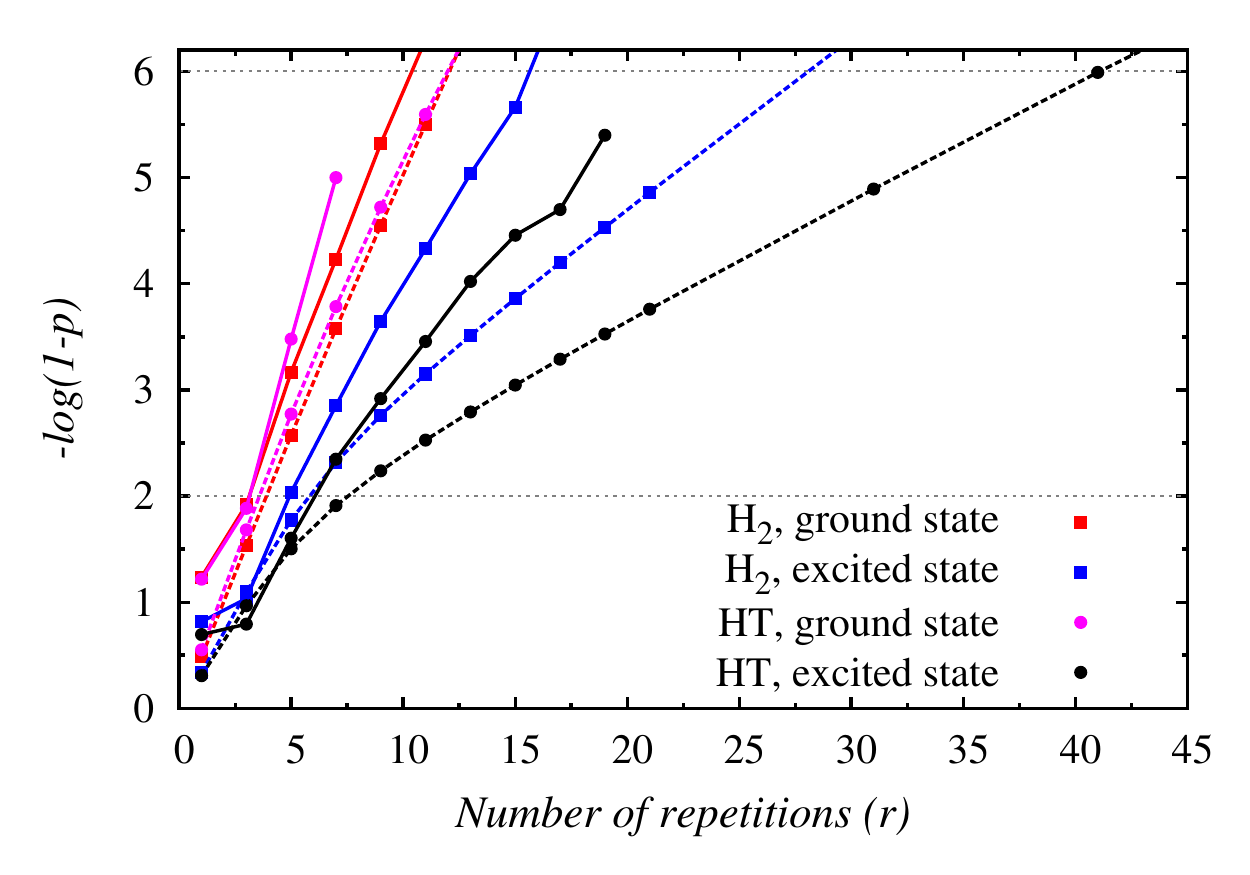}
    \caption{NOMO-TRF/qFCI success probabilities $p$ (IPEA \textbf{A} - solid line, IPEA \textbf{B} - dashed line) for both H$_2$ and HT molecules 
in the basis consisting of $N$ = 6 molecular and $M$ = 10 nuclear orbitals as a function of the number of repetitions ($r$).}
    \label{graf_6_10}
  \end{center}
\end{figure}

\begin{figure}[!ht]
  \begin{center}
    \includegraphics[width=8cm]{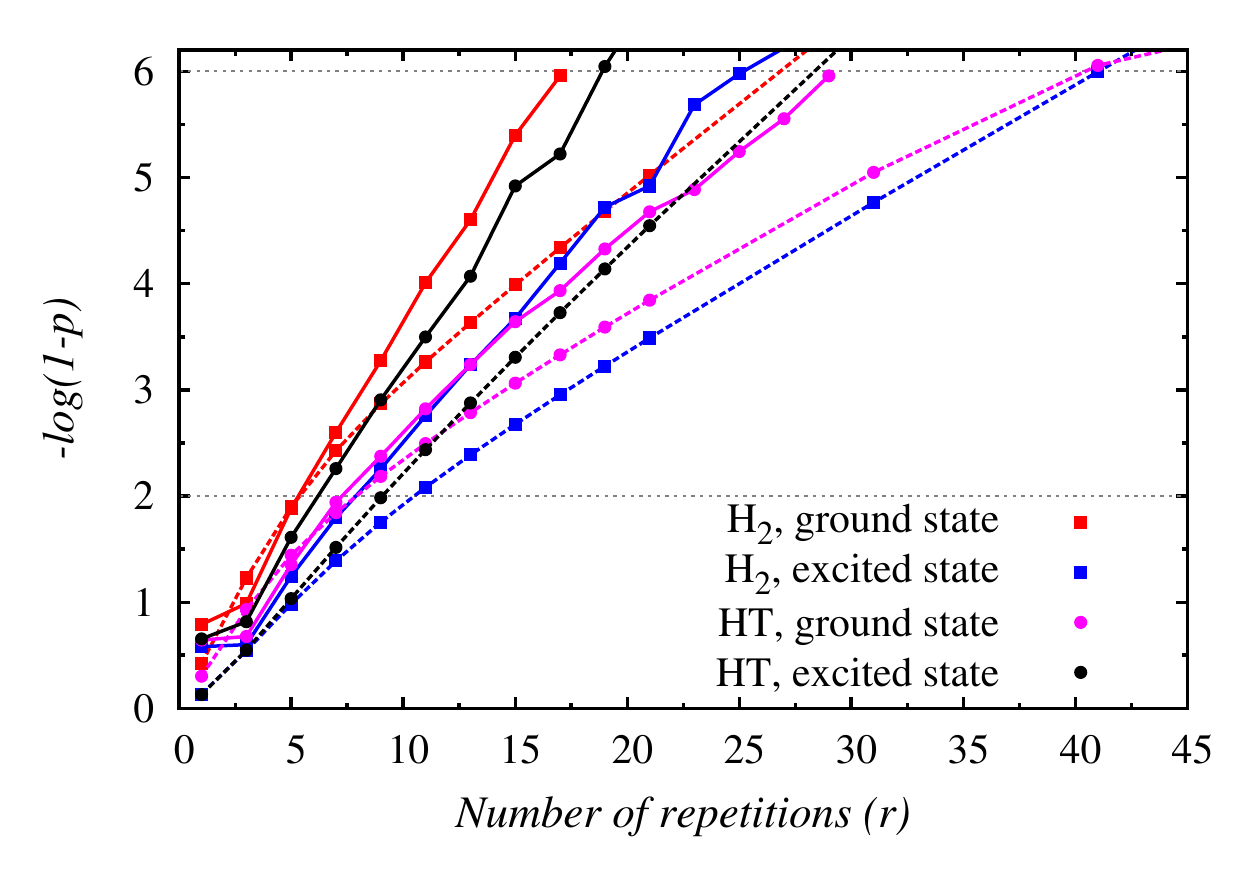}
    \caption{NOMO-TRF/qFCI success probabilities $p$ (IPEA \textbf{A} - solid line, IPEA \textbf{B} - dashed line) for both H$_2$ and HT molecules 
in the basis consisting of $N$ = 9 molecular and $M$ = 23 nuclear orbitals as a function of the number of repetitions ($r$).}
    \label{graf_9_23}
  \end{center}
\end{figure}

The exponential increase of success
 probabilities $p$ with the number of repetitions $r$ (as supposed from the Chernoff bounds \cite{Hagerup_Rub_1990}) is demonstrated in Fig. \ref{graf_6_10} (for the case of basis consisting of $N = 6$
 molecular and $M = 10$ nuclear orbitals) and Fig. \ref{graf_9_23} ($N = 9, M = 23$) by roughly asymptotically linear curves of quantity $f(r) = -\log(1-p)$ as a function of $r$.

%The $r$-coordinates of intersection points between either solid ($x = A$ - for IPEA A) or dashed ($x = B$ - for IPEA B)
% lines and horizontal dashed lines corresponding to $-\log(1-p)$ equal to either $2$ or $6$ are $N_{\rm x,y,z}$ ($y \in \{\rm gs, es \}$ and $z = -\log(1-p)$)
% values shown in Tab. \ref{Tabc2} and Tab. \ref{Tabc4} respectively.

For almost all data points, the success probabilities for
 IPEA \textbf{A} are higher than corresponding success probabilities for IPEA \textbf{B}
% for any number of repetitions 
and the slope of $f(r)$
 curve in asymptotic region is subsequently also higher for IPEA \textbf{A} than for IPEA \textbf{B}.
In most cases, we can also see higher values and slopes of curves corresponding to
ground state when compared to the excited state.
 This correlates with the fact that excited state NOMO-TRF/FCI eigenvector has smaller overlap with its
inital guess than the ground state. In other words, the excited state has stronger multireference character.

%\begin{figure}[ht]
%  \begin{center}
%    \includegraphics[width=12cm]{./grafy/6_23/6_23.eps}
%    \label{graf_6_23}
%    \caption{Success probabilities $p$ (IPEA A - solid line, IPEA B - dashed line) for both ${}^1$H$_2$ and HT molecules for NOMO-TRF/Full-CI in basis consisting of $N$ = 6 molecular and $M$ = 23 nuclear orbitals as a function of the number of repertitions ($r$).}
%  \end{center}
%\end{figure}

\subsection{Discussion}

Results presented in the previous section indicate that for all choices of the FCI active space [defined by the number of molecular ($N$) and nuclear ($M$) orbitals], single determinant initial guesses give sufficiently high success probabilities to be amplified by repetitions ($p > 0.5$). In most cases, success probabilities are higher than $0.75$. The only exception when the success probability is lower than $0.6$ is the case of the H$_2$ excited state and $N = 2$, $M = 6$. This choice of the FCI active space is apparently too small to properly describe the transitional energy anyway.

Apart from the $N = 2$, $M = 6$ active space, results in Tables \ref{Tabc2} and \ref{Tabc4} indicate that at most around $10$ repetitions are sufficient to amplify the success probabilities to $0.99$, and at most $55$ repetitions for amplification to $0.999999$.

For almost all data points, the success probabilities $p$ for
IPEA \textbf{A} are higher than corresponding success probabilities for IPEA \textbf{B} and subsequently slopes of $f(r) = -\log(1-p))$ curves in asymptotic region are also higher.
This could be easily expected as in case of IPEA \textbf{B}, no collapsing of the system and improving the overlap between the actual state of the quantum register and the exact wave function occurs during iterations which is in contrast to IPEA \textbf{A}.

In most cases, we can also note higher values and slopes of curves corresponding to
ground state when compared to the excited state. The only exception is the case of the largest active space used ($N$ = 9, $M$ = 23) for the HT molecule (Fig. \ref{graf_9_23}).
Lower success probabilities for the excited state correlates with the excited state NOMO-TRF/FCI eigenvector having smaller overlap with its
inital guess than the ground state (stronger multireference character of the excited state)

In the exceptional case ($N$ = 9, $M$ = 23 for HT), the overlap for the ground state $S_{\rm gs}$ is
higher than the overlap for the excited state $S_{\rm es}$, but due to the higher phase reminder $\delta$ (see eqs. 8 and 9 in \cite{veis_2010}) for the ground state,
the success probability of non-repeated ($r$ = 1) IPEA \textbf{A} is slightly higher for the excited state than for the ground state.
The phase reminder has the same effect for repeated IPEA success probabilities in both IPEA \textbf{A} and IPEA \textbf{B} cases.

\section{Conclusions}

In this paper we presented  an efficient quantum algorithm for molecular energy computation beyond the Born-Oppenheimer approximation. Our approach is based on the quantum full
configuration interaction method and treats electrons and nuclei on an equal footing, using the nuclear orbital plus molecular orbital (NOMO) method.
We have presented details of the algorithm and demonstrated
its performance by simulations on a classical computer.
For these simulations we have employed relatively small one-particle basis sets and used the compact mapping to keep the number of required qubits manageable.

Two isotopomers of the hydrogen molecule (H$_2$, HT) were chosen as representative examples and calculations
of the lowest rotationless vibrational transition energies were simulated.
For both isotopomers in their ground as well as excited state we have verified that
the single-determinant initial guess yields high enough success probability to be
amplified by repetitions, for both \textbf{A} and \textbf{B} version of IPEA.
At most 10 repetitions were sufficient to amplify the success probability to 0.99 and at most 55 repetitions were necessary to
achieve 0.999999  success probability.
As expected, for most data points the success probability of IPEA \textbf{A} was higher than corresponding IPEA \textbf{B}, due to the improvement of system wave function overlap due to the measurement in the IPEA \textbf{A} procedure.
In most cases, the excited state required more repetitions than the ground state, which is in agreement with our previous experience on electronic-only calculations \cite{veis_2010}.
To conclude, the qFCI approach has been shown to be viable also for simultaneous treatment of electrons and nuclei beyond the
Born-Oppenheimer approximation.

\section*{Appendix}

For derivation of the scaling of the compact boson mapping introduced in Section \ref{section_nomo_qfci},
let us consider the Hamiltonian parameterised by real-valued integrals $V_{pqrs} = V_{rspq}$ and the most demanding 4-index term of a single Trotter step

%Let us consider
%a term in the Trotter expansion of the exponential of the Hamiltonian
%parameterized by the real-valued integrals $V_{pqrs} = V_{rspq}$

\begin{equation}
	\exp \Big(i \tau V_{pqrs} (\hat{a}_p^{\dagger}\hat{a}_q^{\dagger}\hat{a}_s\hat{a}_r + \hat{a}_r^{\dagger}\hat{a}_s^{\dagger}\hat{a}_q\hat{a}_p ) \Big),
	\label{sh12ap}
\end{equation}

\noindent 
with $p$, $q$, $r$, and $s$ mutually different.
We use the lemma

\begin{equation}
\exp(i \tau \Phi \hat{V})=\exp(i \tau \Phi \hat{U} \hat{D} \hat{U}^{\dagger})=\hat{U} \exp(i \tau \Phi \hat{D})\hat{U}^{\dagger},
\label{lemap}
\end{equation}

\noindent
where $\tau$ and $\Phi \equiv V_{pqrs}$ are real numbers, $\hat{V}$ is a hermitian operator and $\hat{D}=\hat{U}^{\dagger} \hat{V} \hat{U}$ is its diagonal form.
It is thus sufficient to find eigenvectors and eigenvalues of the operator 
$\hat{V}=\hat{a}_p^{\dagger}\hat{a}_q^{\dagger}\hat{a}_s\hat{a}_r + \hat{a}_r^{\dagger}\hat{a}_s^{\dagger}\hat{a}_q\hat{a}_p$ 
on a Hilbert space corresponding to the quantum register storing occupation numbers for spinorbitals $p$, $q$, $r$ and $s$. This space is a direct product of two $f_{\rm{max}}(p)+1=f_{\rm{max}}(r)+1$ dimensional and two 
$f_{\rm{max}}(q)+1=f_{\rm{max}}(s)+1$ dimensional spaces. 
In the basis characterized by boson occupation numbers the operator $\hat{V}$ has a block-diagonal structure, since 

\begin{eqnarray}
& \hat{V}|f_p,f_q,f_r,f_s\rangle =  
\nonumber
\\
& = \sqrt{(f_p+1)(f_q+1)f_r f_s} |f_p+1,f_q+1,f_r-1,f_s-1\rangle +
\nonumber
\\
& \sqrt{f_p f_q(f_r+1)(f_s+1)} |f_p-1,f_q-1,f_r+1,f_s+1\rangle.
\nonumber
\\
& 
\end{eqnarray}

\noindent
In the simplest case of $f_{\rm{max}}(p)=f_{\rm{max}}(q)=n$ there are $12(n+1-d)+2-\delta_{n+1,d}$ 
$d$-dimensional blocks ($d=1,2,...,n+1$). In general, the number of diagonal $d$-dimensional blocks ($d=1,2,...,\min(n_1,n_2)+1$) denoted here $p_d$, equals (see the Supplementary Information, Chapter 2.1.)

\begin{eqnarray}
& p_d = -\delta_{z,0}(|n_1-n_2|-1)^2 + 
\nonumber 
\\
& + 2(|n_1-n_2|^2 + 1) + 6 z(|n_1-n_2|+z))
\nonumber
\\
&
\label{pdformula}
\end{eqnarray}

\noindent
where $z=\min(n_1,n_2)+1-d$ and $n_1 = f_{\rm{max}}(p) = f_{\rm{max}}(r)$ and $n_2 = f_{\rm{max}}(q) = f_{\rm{max}}(s)$.
Decomposition of the matrix representation of $\hat{V}$ ($V$) and then subsequently $U$ into blocks leads to a decomposition of $\hat{U}$ from lemma (\ref{lemap}) 
to a direct sum of unitary operators $\hat{U_i}$ acting on each block 

%\begin{equation}
% U = \bigoplus_{i=0}^{K} U_i
%\label{mateq}
%\end{equation}

\begin{equation}
 \hat{U} = \bigoplus_{i=0}^K \hat{U_i}.
\label{opeq} 
\end{equation}

\noindent
Let us define

\begin{equation}
M_{\rm{S},\textit{s}} = \sum_{d=1}^{\min(n_1,n_2)+1} d^s p_d,
\end{equation}
\noindent
where $s$ is an auxiliary non-negative integer. $M_{\rm{S},\textit{s}}$ has a different meaning depending on $s$ value.

For $s = 0$, $M_{\rm{S},\textit{s}}$ equals the total number of subspaces $K = M_{\rm{S},0} = n_1 n_2 (n_1+n_2+1)$ in the decomposition (\ref{opeq}).

The dimension of the quantum register space where operator (\ref{sh12ap}) acts is 
$M_{\rm{S},1} = (n_1+1)^2(n_2+1)^2$.
$M_{\rm{S},1}$ is in fact the minimal possible size of the quantum register for representing operator (\ref{sh12ap}). In the case of qubits, the quantum register dimension will be $2^{2 \lceil \log_2 (n_1 + 1) \rceil + 2 \lceil \log_2 (n_2 + 1) \rceil} \geq (n_1+1)^2(n_2+1)^2 = M_{\rm{S},1}$. Usage of qu-$d$-its for well chosen $d$'s may decrease the ``excess'' dimensions usually padded by unit operator blocks
\noindent
 %, the number of gates needed to implement chain of operators of the right side of (\ref{opeq}) ($s = 2$) 
and the number $C(n_1,n_2)$ of classical precomputing operations needed for the diagonalization of matrix representation (\ref{sh12ap}). 

For $s = 2$, $M_{\rm{S},2} = N_{\rm{g}}(n_1,n_2)$ describes the computational complexity as will be shown in the end of this section. 

The computational cost of classical precomputing operations scales as $C(n_1,n_2) = M_{\rm{S},3}$. 

\begin{figure}[!ht]
 \begin{center}

\hskip 0.5cm
 \includegraphics[width=\textwidth]{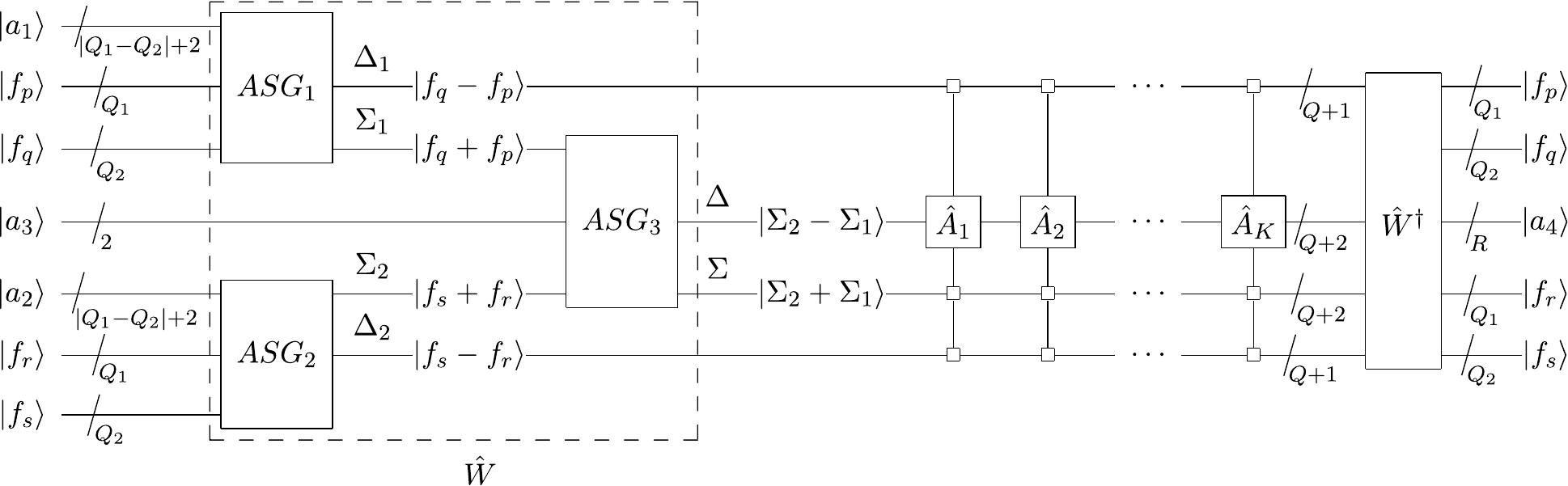}    
 %\includegraphics[width=\textwidth]{Fig4cln02_d.eps} 
 % the last one is an alternative picture - probably better quality

\end{center}
 \caption{The quantum circuit implementing a single Trotter step (boson-boson interaction, see Eq. 
(\ref{sh12ap}), (\ref{lemap}) and (\ref{opeq})). The $\hat{A_i}$ $Q+2$ qubit gates are controlled by $i$-specific $Q+1$, $Q+2$ and $Q+1$ bit sequences, quantum circuit corresponding to them is presented in Fig. \ref{aopqc} with one additional working qubit increasing the number of qubits $\hat{A_i}$ operates on to $Q+3$.
The gate labeled ASG is the Adder-Subtractor Gate realizing the simultaneous addition (sum) and subtraction of two 
binary represented integers. 
The different indices for ASG gates corresponds to different ordering of outputs 
and inputs. The $|Q_1-Q_2|+2$-qubit kets $\ket{a_1}$ and $\ket{a_2}$ are initialy set to zero ($\ket{0}$) and 
correspond to the neccessity to pad the input integer with less bits with $|Q_1-Q_2|$ zeros to match the number of bits of 
the greater input integer prior the particular logical operation, one carry bit is needed for the output integer corresponding
to the sum and one extra bit (sign bit) is needed for the output integer corresponding to the difference. 
Similarly, the ket $\ket{a_3}$ is a 2-qubit initialy set to zero ($\ket{00}$) to provide the most significant bit for 
the sum $\Sigma$ and the sign bit for the difference $\Delta$. The ket $\ket{a_4}$ is a $R = 2|Q_1-Q_2|+6$ register for the recovery of all working qubits.
}
 \label{uiimplement}
\end{figure}

Based on the above approach, the exponential (\ref{sh12ap}) can be decomposed into blocks,

\begin{equation}
\exp(i \tau \Phi \hat{V}) = \bigoplus_{i=0}^K \hat{A_i},
\label{adef}
\end{equation}

\begin{equation}
\hat{A_i} = \hat{U_i} \exp(i \tau \Phi \hat{D}_i) \hat{U_i}^{\dagger},
\label{aop}
\end{equation}

\noindent where $\hat{D}_i$ is a diagonal gate with only $d_i$ non-zero elements. Gates $\hat{A_i}$ in the circuit correspond to the product (\ref{aop}) and are further described in Fig. \ref{aopqc}. Before and after the sequence of $\hat{A_i}$ gates is applied, the transformation

\begin{eqnarray}
& |f_p,f_q,f_r,f_s\rangle \mapsto |\Sigma_1,\Delta_1,\Delta_2,\Sigma_2\rangle \mapsto
\nonumber
\\
& \mapsto |\Delta,\Delta_1,\Delta_2,\Sigma\rangle,
\label{rear}
\end{eqnarray}

\noindent and its inverse have to be applied as shown in Fig. \ref{uiimplement} (The transformation (\ref{rear}) is realized by the subcircuit in the dashed box (Fig. \ref{uiimplement}, the gate $\hat{W}$).

The non-negative integers $f_p$, $f_q$, $f_r$ and $f_s$ are occupation numbers, 
their upper-bounds are $f_p,f_r\leq n_1$ and $f_q,f_s\leq n_2$ 
(and corresponding register sizes in qubits $Q_1 = \lceil\text{log}_2(n_1+1)\rceil$ and
$Q_2 = \lceil\text{log}_2(n_2+1)\rceil$). In formula (\ref{rear}) the first 
transformation produces the sum and difference of occupation number pairs ($f_p$, $f_q$) 
and ($f_r$, $f_s$),  

\begin{equation}
\Sigma_1 = f_q + f_p,
\label{transf01}
\end{equation}

\begin{equation}
\Delta_1 = f_q - f_p,
\label{transf02}
\end{equation}

\begin{equation}
\Sigma_2 = f_s + f_r,
\label{transf03}
\end{equation}

\begin{equation}
\Delta_2 = f_s - f_r,
\label{transf04}
\end{equation}

\noindent
and the second transformation in (\ref{rear}) produces sum and difference of the first and third registers denoted $\Sigma_1$ and $\Sigma_2$ respectively,

\begin{equation}
\Sigma = \Sigma_2 + \Sigma_1,
\label{transf05}
\end{equation}

\begin{equation}
\Delta = \Sigma_2 - \Sigma_1.
\label{transf06}
\end{equation}

The whole Trotter step (\ref{sh12ap}) is represented by a quantum circuit model in Fig. \ref{uiimplement} and the particular implementation of the ASG gate is discussed in detail in Chapter 2.3 of the Supplementary Information.
Note that the ASG gate can be realized by $\mathcal{O}(Q)$ elementary gates and (depending on a particular realization) with 0 to $Q$ working qubits.
The existence of the ASG gate is obvious from the fact that from the combination of the sum and difference of the
two input integers, the input integers can be unambiqously deduced. The ASG gate outputs for the ancilla registers $\ket{a_i}$ different
from zeros are irrelevant.

%\begin{widetext}

\clearpage
\newpage

\newcommand{\hhat}[1]{\mbox{$\hat{#1}$}}

\begin{figure}[!ht]
 \begin{center}
 \hskip 1.8cm
 \mbox{
   \Qcircuit @C=1.5em @R=1em {
     \lstick{\ket{(\Delta)_1}} & \multigate{8}{ADD(-\Delta_{0,i})} & \ctrlo{1} & \ctrlo{1} & \ctrlo{1} & \multigate{8}{ADD(\Delta_{0,i})} & \push{\ket{(\Delta)_1}} \qw \\
     \lstick{\ket{(\Delta)_2}} & \ghost{ADD(-\Delta_{0,i})} & \ctrlo{1} & \ctrlo{1} & \ctrlo{1} & \ghost{ADD(\Delta_{0,i})} & \push{\ket{(\Delta)_2}} \qw \\
     \lstick{\ket{(\Delta)_3}} & \ghost{ADD(-\Delta_{0,i})} & \multigate{2}{\hhat{U'_i}} & \multigate{2}{\exp(i \tau \Phi \hhat{D'_i})} & \multigate{2}{\hhat{U'^{\dagger}_i}} & \ghost{ADD(\Delta_{0,i})} & \push{\ket{(\Delta)_3}} \qw \\
     \lstick{\enspace \vdots \enspace} & \ghost{ADD(-\Delta_{0,i})} & \ghost{\hhat{U'_i}} & \ghost{\exp(i \tau \Phi \hhat{D'_i})} & \ghost{\hhat{U'^{\dagger}_i}} & \ghost{ADD(\Delta_{0,i})} & \push{\enspace \vdots \enspace} \qw \\
     \lstick{\ket{(\Delta)_{\lceil \log_2 d_i \rceil + 2}}} & \ghost{ADD(-\Delta_{0,i})} & \ghost{\hhat{U'_i}} & \ghost{\exp(i \tau \Phi \hhat{D'_i})} & \ghost{\hhat{U'^{\dagger}_i}} & \ghost{ADD(\Delta_{0,i})} & \push{\ket{(\Delta)_{\lceil \log_2 d_i \rceil + 2}}} \qw \\
     \lstick{\ket{(\Delta)_{\lceil \log_2 d_i \rceil + 3}}} & \ghost{ADD(-\Delta_{0,i})} & \ctrlo{-1} & \ctrlo{-1} & \ctrlo{-1} & \ghost{ADD(\Delta_{0,i})} & \push{\ket{(\Delta)_{\lceil \log_2 d_i \rceil + 3}}} \qw \\
     \lstick{\enspace \vdots \enspace} & \ghost{ADD(-\Delta_{0,i})} & \ctrlo{-1} & \ctrlo{-1} & \ctrlo{-1} & \ghost{ADD(\Delta_{0,i})} & \push{\enspace \vdots \enspace} \qw \\
     \lstick{\ket{(\Delta)_{Q+2}}} & \ghost{ADD(-\Delta_{0,i})} & \ctrlo{-1} & \ctrlo{-1} & \ctrlo{-1} & \ghost{ADD(\Delta_{0,i})} & \push{\ket{(\Delta)_{Q+2}}} \qw \\
     \lstick{\ket{(\Delta)_{Q+3}}} & \ghost{ADD(-\Delta_{0,i})} & \ctrlo{-1} & \ctrlo{-1} & \ctrlo{-1} & \ghost{ADD(\Delta_{0,i})} & \push{\ket{(\Delta)_{Q+3}}} \qw \rstick{\hhat{A'_i}} \gategroup{3}{3}{5}{5}{.7em}{--} \\
   }
}
 \end{center}
 \caption{The quantum circuit representing the action of the gate $\hat{A}_i$ (which occures in Fig. \ref{uiimplement} and equations (\ref{adef}) and (\ref{aop})) on its 
target quantum register $\Delta$ padded by one ancilla qubit $(\Delta)_{Q+3}$ needed for ADD gate operation.
The Q+3 qubit gate ADD($a$) adds a constant integer $a = \pm \Delta_{0,i}$ to the input quantum register producing the output quantum register.
The ADD gate is based on the generalized $\Phi$ ADD gate (operating in $\mathcal{O}(1)$ time, see the Fig. 4 
on the page 654 of \cite{pavlidis_gizopoulos_2014}) inserted between the forward and backward Quantum Fourier Transforms 
(operating in $\mathcal{O}(Q \log Q)$ time each). The value $\Delta_{0,i}$ corresponding to the $i$-th subspace (in the decomposition (\ref{opeq})) is the smallest value of $\Delta$ corresponding to any vector in that subspace and is 
to be precomputed classicaly for each $i$ in $\{1, 2, \ldots, K \}$. The primes correspond to the fact that the operators now act on different subspaces.
In the special case of one dimensional subspaces ($d_i = 1$), $\hat{U'_i}$, $\exp(i \tau \Phi \hat{D'_i})$ and $\hat{U'^{\dagger}_i}$ 
gates have 0 qubit 
target register and are therefore equal to multiply controlled phase-shifts. Since the corresponding phase is $1 = \exp(0)$ for all
of them, they can be omitted from the quantum circuit. 
}
\label{aopqc}
\end{figure}
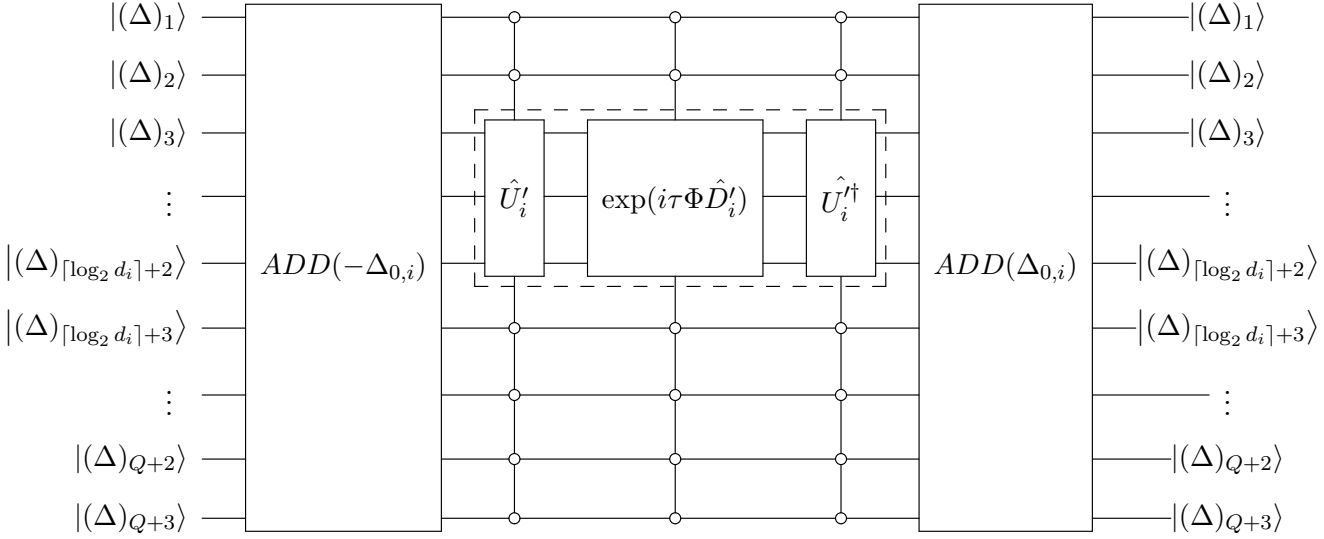

\clearpage
\newpage

%\end{widetext}

\noindent
The transformation (\ref{rear}) needs $2|Q_1-Q_2|+4$ ancilliary qubits for the first part 
and another 2 ancilliary qubits for the second part. Therefore the first and the 
last quantum registers (storing the values of $\Delta$ and $\Sigma$ respectively) 
have a size $\text{max}(Q_1,Q_2) + 2$ qubits while the middle quantum registers 
(storing the values of $\Delta_1$ and $\Delta_2$) of $\text{max}(Q_1,Q_2) + 1$ qubits. 
In the following text, $Q \equiv \text{max} (Q_1,Q_2)$. 
The number of single qubit gates and CNOTs for the transformation (\ref{rear}) scales as 
$\mathcal{O} (Q)$ if the algorithm presented in \cite{vedral_1995} is used (with another $Q+1$ working qubits) or as $\mathcal{O} (Q \text{log} Q)$ (but with no need for further working qubits) 
if the algorithm exploiting QFT \cite{florio_picca}, \cite{ruiz-perez_garcia-escartin_2014} is employed. Each $\hat{U_i}$ quantum gate acts 
on the target register which is subregister of the first register in the last part of (\ref{rear}), 
storing the value of $\Delta$ and is multiply controlled by other qubits 
representing the ket $|\Delta,\Delta_1,\Delta_2,\Sigma\rangle$. Multiply controlled quantum gates can be
decomposed into the bare quantum gate acting on the target register and either $\mathcal{O} (Q)$ 
(1-qubit and CNOT) gate cost with using $3Q+2$ working ancilliary qubits or $\mathcal{O} (Q^2)$ 
gate cost without any working ancilliary qubits 
(for the algorithm see \cite{nielsen_chuang} on pages 183 and 193 respectively, the variant with ancilliary qubits is also mentioned in \cite{agaian_klappenecker}). The action of $\hat{A}_i$ gate from Fig. \ref{uiimplement} and equations (\ref{adef}) and (\ref{aop}) is described 
in the quantum circuit in Fig. \ref{aopqc}.
The bare quantum gate from Fig. \ref{aopqc}, $\hat{U'_i}$, acting on the target register ($2^{\lceil\text{log}(d)\rceil}$-dimensional subspace) can be decomposed 
(via Quantum Shanon Decomposition (QSD), \cite{Shende_2006})
 into 
$\mathcal{O} (d^2)$ elementary quantum gates.  
Neglecting the contribution 
of gates acting on the controlling register which scales as $\mathcal{O} (M_{\rm{S},0} Q^2) = \mathcal{O} (x^2 y \text{log}^{2} (x+1))$ in the 
worst case, the formula for
$N_{\mathrm{gates}}(n,m)$ (\ref{ngat}) is derived as $M_{\rm{S},2}$.  The diagonal bare quantum gate $\exp(i \tau \Phi \hat{D'_i})$ can be decomposed into the 1-qubit gates and CNOTs at most with the same effort as $\hat{U'_i}$ as further discussed in the Supplementary Information (Chapter 2.4.).
It is important to note, that the classical pre-processing - diagonalization of $\hat{V}$ and QSD of the corresponding $\hat{U_i}$ 
operators needs to be done just once (for e.g. $p = 1, q = 2, r = 3, s = 4$) before the quantum algorithm is started, then for each elementary Trotter term (\ref{sh12ap}) the seqence of quantum gates differs just by addressing different quantum registers (no longer $p = 1, q = 2, r = 3, s = 4$) and by different value of $\Phi = V_{pqrs}$ as term in phase-parameter in $\exp(i \tau \Phi \hat{D})$ 
diagonal operator from lemma (\ref{lemap}).

\section*{Acknowledgement}
This work has been supported by the Grant Agency of the Czech Republic - GA\v{C}R (203/08/0626)
and by the Charles University project ``Student research in biophysics and chemical physics'' (SVV 260214).
\bibliographystyle{h-physrev}
\bibliography{kvantove_pocitace}

\end{document}